\documentclass[11pt]{article}

\def\be{\begin{equation}}
\def\ee{\end{equation}}
\def\bea{\begin{eqnarray}}
\def\eea{\end{eqnarray}}

\def\ell{l}
\usepackage{enumerate}

\title{\bf{Near-horizon geometries of supersymmetric \\ $AdS_5$ black holes}}
\author{Hari K. Kunduri \\School of Physics and Astronomy, University of Nottingham, NG7 2RD, UK \\h.k.kunduri@nottingham.ac.uk \\ \\ James Lucietti \\ Centre for Particle Theory, Department of Mathematical Sciences, \\ University of Durham, South Road, Durham, DH1 3LE, UK\\ james.lucietti@durham.ac.uk \\   }
\date{28 August, 2007 ({\small DCPT-07/49})}

\begin{document}

\maketitle

\begin{abstract}
We provide a classification of near-horizon geometries of
supersymmetric, asymptotically anti-de Sitter, black holes of
five-dimensional $U(1)^3$-gauged supergravity which admit two
rotational symmetries. We find three possibilities: a topologically
spherical horizon, an $S^1 \times S^2$ horizon and a toroidal
horizon. The near-horizon geometry of the topologically spherical
case turns out to be that of the most general known supersymmetric, asymptotically anti-de Sitter, black hole of $U(1)^3$-gauged supergravity. The other two cases have constant scalars and only exist in particular
regions of this moduli space -- in particular they do not exist
within minimal gauged supergravity. We also find a solution
corresponding to the near-horizon geometry of a three-charge
supersymmetric black ring held in equilibrium by a conical
singularity; when lifted to type IIB supergravity this solution can be made regular, resulting in a discrete family of
warped $AdS_3$ geometries. Analogous
results are presented in $U(1)^n$ gauged supergravity.
\end{abstract}

\section{Introduction}
Supersymmetric, asymptotically $AdS_5 \times S^5$, black holes have
only been known for a few years~\cite{GR1,GR2,Chongetal,KLR}.  Their
relevance stems from the AdS/CFT correspondence~\cite{adscft}, which
implies such black holes should be dual to 1/16 BPS states of
$\mathcal{N}=4$ $SU(N)$ Yang-Mills theory on ${R} \times S^3$. Such
states are, generically, specified by five quantum numbers: the
$SO(4)$ spins $J_1,J_2$ and the $SO(6)$ R-charges $Q_1,Q_2,Q_3$.
However, the most general {\it known} black hole of this kind has a
constraint relating these leaving only four independent conserved
charges~\cite{KLR}. Indeed, this is the most general known
regular\footnote{Here we allow for solutions regular on and outside
an event horizon, which typically possess a singularity behind the
horizon and thus are not globally regular.}, asymptotically $AdS_5
\times S^5$, 1/16 BPS, solution of type IIB supergravity. Therefore,
before attempting to derive the Hawking-Bekenstein entropy from a
microscopic counting in the field theory, this discrepancy in the
number of 1/16 BPS states between the two dual pictures needs to be
understood~\cite{KMMR}.

There are  three ways this potential contradiction could get
resolved. One possibility is that we already know the most general black
hole and due to finite coupling effects in the CFT only a four parameter subset
contribute to the $O(N^2)$ entropy at large $N$ (see e.g.~\cite{berkooz}). A second
possibility is that there is a sufficient number of asymptotically
$AdS_5 \times S^5$, 1/16 BPS, solitons\footnote{By this we mean
smooth, strictly stationary solutions -- therefore horizonless.} to
account for the missing states, although we note that no examples of
such solutions are known\footnote{These would be analogues of the
1/2 BPS LLM geometries~\cite{LLM}. However, even the 1/4 and 1/8 BPS cases are
far from being understood~\cite{Chen}. We should also note that it is not clear that for given conserved charges black holes and solitons should be counted on the same footing.}. The final possibility, which we shall
explore in this paper, is that there is a more general family of
black holes. This latter possibility can be realised in a number of
ways, raising the following questions:
\begin{enumerate}
\item The known solutions are all solutions to five dimensional gauged supergravity, which is a consistent truncation of type IIB supergravity on $S^5$ \footnote{It is only known how to perform the reduction from the subsector of IIB which keeps the metric and self-dual five form~\cite{reduction}.}; do asymptotically $AdS_5 \times S^5$ black holes exist within type IIB that are truly ten dimensional? -- i.e. depend on the higher spherical harmonics on $S^5$.
\item The known solutions are solutions to a truncation of the maximal 5d $SO(6)$-gauged supergravity where one only keeps the maximal abelian subgroup $U(1)^3$ (with two scalars). Are there 5d black holes with extra scalars and/or non-abelian gauge fields turned on?
\item The known class of black holes possess topologically spherical horizons (in 5d language); are there black holes with more exotic horizon topology, such as black rings?
\item Is there a more general black hole in the 5d $U(1)^3$ gauged supergravity than the currently known one?
\end{enumerate}

In this paper we will only be concerned with black hole solutions of
5d $U(1)^3$ gauged supergravity\footnote{By this we mean minimal
gauged supergravity coupled to two abelian vector multiplets.} and
therefore can only address questions 4 and 3 (partially). Black hole
uniqueness theorems have not been established for asymptotically AdS
spacetimes, even in four dimensions. Further, it is known that these
theorems fail for asymptotically flat spacetimes in five dimensions~\cite{vacring}.
It is therefore unclear how many $AdS_5$ black hole solutions one
should expect to exist for given conserved charges\footnote{In
particular it is not clear whether they should be uniquely specified
by their conserved charges, even for the case of spherical topology.
Indeed, if one assumes this is the case, then a non-extremal black
hole would be parameterised by 6 conserved charges:
$M,J_1,J_2,Q_1,Q_2,Q_3$. In general the BPS limit will reduce the
number of parameters of a black hole by two: this is because
extremality and supersymmetry do not lead to the same constraint.
This leaves one with a 4 parameter black hole, presumably the one
which is already known~\cite{KLR}.}. However, we are concerned with
supersymmetric solutions. There exist systematic methods that help
one construct such solutions~\cite{GG,GR2}. Unfortunately, guesswork is still
required (i.e. choice of a K\"ahler metric on the base manifold), thus leaving the task of a full classification of
supersymmetric $AdS_5$ black holes out of reach for the moment.

Since supersymmetric black holes are extremal they admit a
near-horizon limit. This limit allows one to focus on a region in
the neighbourhood of the horizon in such a way that the limit is a
solution to the same theory the black hole is. Thus a more modest
goal presents itself: a classification of the near-horizon
geometries of asymptotically $AdS_5$ black holes.

Recently, we performed such an analysis within minimal gauged
supergravity~\cite{KLR2} under the assumption that the black hole
admits two rotational symmetries.  It was
shown the near-horizon geometry of an asymptotically $AdS_5$
supersymmetric black hole, admitting two rotational symmetries must
have a horizon of spherical topology and is given by the
near-horizon limit of the black holes of~\cite{Chongetal}. In
particular this ruled out the existence of supersymmetric black
rings with these symmetries.

The purpose of this paper is to generalise this analysis to the
$U(1)^3$ gauged supergravity. Qualitatively, our results differ to
those we found in minimal supergravity. In $U(1)^3$ gauged
supergravity we find that the most general topologically spherical
supersymmetric black hole, with two rotational symmetries, is the
near-horizon geometry of the solution found in~\cite{KLR} (this
result is analogous to the one in minimal supergravity). However, we
find that other topologies with these symmetries are also allowed:
namely $S^1 \times S^2$ and $T^3$ . The corresponding near-horizon
solutions are $AdS_3 \times S^2$ and $AdS_3 \times T^2$ respectively
and must have constant scalars\footnote{Such product geometries in
$U(1)^3$-gauged supergravity have in fact been noticed
before~\cite{Cacciatori}.}. These solutions only exist in a
particular region of the scalar moduli space which does not include the minimal
theory. Further, the $S^1 \times S^2$ case is only a three parameter family of solutions, whereas the near-horizon of the supersymmetric black ring in ungauged supergravity~\cite{multibpsring} is a four parameter family.
Nevertheless this means we are not able to rule out the existence of
supersymmetric anti-de Sitter black rings in $U(1)^3$-gauged
supergravity. However, we should emphasise, that the existence of a near-horizon geometry does \emph{not} imply that a corresponding black hole solution with prescribed asymptotics (i.e. in our case global $AdS_{5}$) exists.

The above three cases are the only possible regular near-horizon geometries with two rotational symmetries.  However as in our analysis of minimal supergravity~\cite{KLR2} we also find
a solution describing the near horizon of an unbalanced black ring
supported by a conical singularity. It consists of a warped product
of $AdS_3$ and a singular $S^2$ and reduces to the near-horizon
of the black ring in ungauged supergravity. When lifted to type IIB
supergravity we show that this solution can be made regular, resulting in a discrete set of regular
warped $AdS_3$ geometries.

This paper is organized as follows. In the Section 2, we summarise
the main results of our analysis for $U(1)^3$ supergravity. The
remainder of the paper will be dedicated to the derivation of these
results. We found it convenient to work in the more general $U(1)^n$
gauged supergravity. In Section 3 we derive the conditions imposed
by supersymmetry and the equations of motion in the near-horizon
limit. In section 4 we determine all the possible near-horizon geometries of a supersymmetric AdS black hole with $U(1)^2$
spatial isometry (i.e. two rotational symmetries) and perform a detailed global analysis of these geometries.
We conclude with a Discussion. Some details of the
analysis are given in the Appendix.

\section{Main results}
In this section we explicitly state the main results of this paper.
This section is intended to be a self-contained account without
derivations.

The action for the bosonic sector of five-dimensional $U(1)^3$-gauged supergravity is: \bea
\label{u1cubed}
S &=& \frac{1}{16 \pi G} \int d^5x \left( \sqrt{g} \left[ R +4g^2 \sum_{I=1}^3 (X^I)^{-1}- \frac{1}{2} \sum_{I=1}^3 (X^I)^{-2} (\partial X^I)^2 - \frac{1}{4}\sum_{I=1}^3 (X^I)^{-2} F^I \cdot F^I \right] \right.\nonumber \\
&& \qquad \qquad \qquad \qquad  \left. -F^1 \wedge F^2 \wedge A^3
\right) \eea where $F^I=dA^I$ and the scalars satisfy the constraint
$X^1X^2X^3=1$ ($g$ is the gauge coupling). Note that minimal gauged
supergravity is the following truncation of the above theory:
$X^I=1$ and $A^I=A$. Consider a supersymmetric, asymptotically
$AdS$, black hole solution of the above theory with isometry group
$R \times U(1) \times
U(1)$ (this corresponds to time translational symmetries and spatial rotational symmetries). Spatial sections of the horizon of such a black hole must be: topologically spherical, $S^1 \times S^2$ or toroidal. Below we list the most general near-horizon solutions corresponding to these cases.

\paragraph{Topologically spherical horizon:}The near-horizon solution in this case is always non-static.
The main assumption of our analysis is that the solutions possess $U(1)^2$ rotational symmetry. However, as is typical of rotating solutions in 5d, we find there is a special case for which one has a symmetry enhancement of the rotational group to $SU(2)\times U(1)$. Generically though we find one only\footnote{In contrast, in the ungauged theory one \emph{must} have $SU(2)\times U(1)$ rotational symmetry (this is the near-horizon of the BMPV black hole) \cite{HDBH,Gut}.} has the $U(1)^2$. As a byproduct of our analysis we have found efficient coordinate systems which describe these two cases separately.

The more symmetric and hence simplest case can be written as \bea ds_{NH}^2 &=&
-(\Delta^2+ g^2X^2)r^2 dv^2 +2dvdr +\left[
\frac{\Delta}{\Delta^2+g^2\lambda}(d\phi +\cos\theta d\psi)- gXrdv
\right]^2 \nonumber \\ &+& \frac{1}{\Delta^2 +g^2\lambda}( d\theta^2
+\sin^2\theta d\psi^2) \\ A^I &=& \Delta X^I rdv - \frac{gX^I(X -2X^I)}{\Delta^2+g^2\lambda}(d\phi+\cos\theta d\psi) \eea where the
scalars $X^I$ are constant and we have defined the constants: \bea \label{X}
X& \equiv & X^1+X^2+X^3
\\ \label{lambda}\lambda & \equiv & (X^1)^2 +(X^2)^2+(X^3)^2 -2X^1X^2-2X^1X^3-2X^2X^3.
\eea The solution is parameterized by the constants $(X^I,\Delta)$
subject to the constraints $\Delta>0$ and $\Delta^2+g^2 \lambda>0$
(and of course $X^1X^2X^3=1$) and is therefore a three parameter family. The coordinate ranges are: $0 \leq
\psi < 2\pi$, $0 \leq \phi <4\pi$, $0 \leq \theta \leq \pi$. The horizon is located
at $r=0$ and on spatial sections is a homogeneously squashed $S^3$.
The near-horizon geometry in this case is a homogeneous space and is
a fibration of $AdS_2$ over the homogeneously squashed $S^3$ with
symmetry group $SO(2,1) \times SU(2) \times U(1)$. This solution turns out to be the near-horizon limit of the black holes found in~\cite{GR2}\footnote{The near-horizon geometries of these black holes have been previously studied in~\cite{davis,sinha1,sinha2}.}, which are characterised by $J_1=J_2$ and are special cases of the more general solution of~\cite{KLR}.

The generic case is more complicated. In this case the near-horizon solution can be
written as:
 \bea
 \label{gennonstatic}
ds^2_{NH} &=& H(x)^{1/3} \left[ -C^2R^2dv^2 +2dvdR + \left( C^2-\frac{\Delta_0^2}{H(x)} \right) \left( dx^1+ \omega(x) dx^2 +Rdv \right)^2 \right] \nonumber \\ &+& \frac{H(x)^{1/3}dx^2}{4g^2 P(x)} + \frac{4g^2 H(x)^{1/3} P(x) (dx^2)^2}{ C^2 H(x)-\Delta_0^2}, \\
X^I &=& \frac{H(x)^{1/3}}{x+ 3K_I}, \qquad  A^I=
\frac{X^I}{H(x)^{1/3}} \left[ \Delta_0(Rdv +dx^1)+ (x-\alpha_0)dx^2
\right] \eea where $C^2,\alpha_0,\Delta_0,K_I$ are constants such
that $C,\Delta_0>0$ and $K_1+K_2+K_3=0$ and the functions are
defined by \bea H(x) &=& (x+3K_1)(x+3K_2)(x+3K_3), \qquad  P(x)=
H(x) - \frac{C^2}{4g^2}(x-\alpha_0)^2 - \frac{\Delta_0^2}{C^2}, \\
\omega(x) &=& \frac{\Delta_0(\alpha_0 -x)}{C^2 H(x)-\Delta_0^2}.
\eea
Due to a scaling symmetry of the solution it turns out one parameter is trivial (for instance one can set $C$ to any desired value); therefore this is a 4-parameter family of solutions.
The coordinate ranges are $x_1 \leq x \leq x_2$ where
$0<x_1<x_2<x_3$ are the three roots of the cubic $P(x)$, such that
$x_1+3K_I>0$, and the coordinates defined by $\partial /\partial
\phi_i \propto \omega(x_{i})\partial / \partial x^1 - \partial /
\partial x^2$ ($i=1,2$) are $2\pi$-periodic. The horizon is located
at $R=0$ and spatial sections of this must possess $S^3$ topology
with $\partial /\partial \phi_1$ vanishing at the pole $x=x_1$ and
$\partial / \partial \phi_2$ vanishing at the pole $x=x_2$ -- these
are the generators of the $U(1)^2$ rotational symmetries. In this
case the near-horizon solution is a fibration of $AdS_2$ over a
(non-homogeneously) squashed $S^3$ with an $SO(2,1) \times U(1)^2$
symmetry group and is cohomogeneity-1. It turns out it is 1/2 BPS within $U(1)^3$-gauged supergravity\footnote{This follows from the fact that it is related by an analytic continuation (of the form considered in~\cite{KLR3}) to the Sabra-Klemm time machines~\cite{SK}.}. This solution turns out to be the near-horizon limit of the black holes found in~\cite{KLR} which have $J_1 \neq J_2$.

Therefore the most general near-horizon solution with a topologically spherical horizon and two rotational symmetries, does in fact turn out to be the near-horizon limit of the 4-parameter black holes of~\cite{KLR}. The coordinates of that paper, while not allowing such compact expressions as above, do in fact cover both of the above cases.

The $SO(2,1)$ symmetry of the above near-horizon solutions is guaranteed from the general results of~\cite{KLR3}.

\paragraph{ $S^1 \times S^2$ horizon:}The near-horizon solution is static
and takes the form: \bea \label{NHring} ds^2_{NH} &=& 2dvdr - 2gXrdvdz +
dz^2+ \frac{1}{g^2 \lambda} ( d\theta^2 + \sin^2\theta d\psi^2), \\
A^I &=& -\frac{1}{g \lambda}X^I(X -2X^I)\cos\theta d\psi \nonumber \eea where the
scalars $X^I$ are constants and $X,\lambda$ are the constants
defined in~(\ref{X}) and (\ref{lambda}). The solution is parameterised by the constants $(L,X^I)$, where $L$ is the period of $z$ (which can be arbitrary), subject to the constraints $\lambda>0$ (and of course $X^1X^2X^3=1$) and therefore is a three parameter family. Note that the constraint $\lambda>0$ can be satisfied by $\sqrt{X^1}+\sqrt{X^2}
< \sqrt{X^3}$ (or permutations of $123$) for example. Also observe
that this solution does not exist in minimal gauged supergravity
because in this case $X^I=1$ and therefore $\lambda=-3$. The near-horizon geometry in this case is $AdS_3 \times
S^2$. The horizon is located at $r=0$ and spatial sections are $S^1
\times S^2$ equipped with a direct product
metric with a round $S^2$. This solution is 1/2 BPS~\cite{Cacciatori}.

\paragraph{Toroidal  horizon:} The near-horizon solution is static and given by \bea
ds_{NH}^2 = 2dvdr -2gXrdvdz +dz^2 +dx^2 +dy^2, \qquad A^I = \frac{gX^I}{2}(X - 2X^I)(xdy - ydx) \eea where the scalars $X^I$ are constants which
satisfy $\lambda=0$, and $X,\lambda$ are defined by (\ref{X}) and (\ref{lambda}). This can be achieved by taking
$\sqrt{X^1}+\sqrt{X^2}= \sqrt{X^3}$ (or permutations of $123$) for
example. This solution is not allowed in minimal gauged
supergravity. The near-horizon geometry in this case is $AdS_3 \times T^2$. The
horizon is at $r=0$ and spatial sections are of toroidal
topology equipped with a flat metric. This solution is 1/2 BPS~\cite{Cacciatori}. \\

The above three cases are the only possible { \it regular }
near-horizon geometries with two rotational symmetries. Therefore, in particular, if an anti-de Sitter black ring with two rotational
symmetries exists in this theory, it \emph{must} have the near-horizon geometry
given by~(\ref{NHring}). We will consider this possibility in the Discussion.

Curiously, though, within
our analysis we did find another $S^1 \times S^2$ case, where the
$S^2$ necessarily possesses a conical singularity at one of its
poles. The
solution in this case is locally given by the $\Delta_0 \to 0$
limit of the topologically spherical case (\ref{gennonstatic}) and is also a four parameter family (despite losing the parameter $\Delta_0$ one aquires a fourth parameter from the period of the $S^1$). Note that the $g\to 0$ limit of this solution reduces to the
non-singular four-parameter near-horizon geometry of the asymptotically flat
supersymmetric
black ring of ungauged supergravity, $AdS_3 \times S^2$~\cite{multibpsring}. This could therefore correspond to the near-horizon limit of an
{\it unbalanced} supersymmetric anti-de Sitter black ring.
 For $g>0$ this singular near-horizon geometry can in fact be made regular when oxidised to IIB supergravity; however the resulting geometry can no longer be viewed as a solution of 5d supergravity.
\\

This completes the list of all possible near-horizon limits of supersymmetric $AdS_5$ black holes with symmetry $R \times U(1) \times U(1)$,
in $U(1)^3$-gauged supergravity. In the subsequent sections we perform a
systematic analysis where we prove these are the only possibilities. We actually found it convenient to work in the more general $U(1)^n$
supergravity and thus we have obtained analogues of the above
results in this theory.

\section{Supersymmetric near-horizon geometries}

\subsection{Gauged supergravity} \label{sec:theory}

We shall consider the theory of five dimensional ${\cal N}=1$ gauged supergravity coupled to $n-1$ abelian vector multiplets following the conventions of~\cite{GR2}. The bosonic sector of this theory consists of the graviton, $n$ vectors $A^I$ and $n-1$ real scalars. The latter can be replaced with $n$ real scalars $X^I$ subject to a constraint
\be
 \frac{1}{6}C_{IJK} X^I X^J X^K = 1,
\ee
where $C_{IJK}$ are a set of real constants symmetric under permutations of $(IJK)$. Indices $I,J,K, \ldots$ run from $1$ to $n$. It is convenient to define
\be
\label{eqn:Xconstr}
 X_I \equiv \frac{1}{6} C_{IJK} X^J X^K.
\ee
The action is\footnote{We use a positive signature metric.}
\be
 S = \frac{1}{16 \pi G} \int \left( R_5 \star 1  - Q_{IJ} F^I \wedge \star F^J - Q_{IJ}     dX^I \wedge \star dX^J - \frac{1}{6} C_{IJK} F^I \wedge F^J \wedge A^K  + 2\chi^2 {\cal V} \star 1 \right),
\ee
where $F^I \equiv dA^I$ and
\be
 Q_{IJ} \equiv \frac{9}{2} X_I X_J - \frac{1}{2} C_{IJK} X^K.
\ee For simplicity, we shall assume that the scalars parameterize a
symmetric space, which is equivalent to the condition \be
 C_{IJK} C_{J' (LM} C_{PQ) K'} \delta^{JJ'} \delta^{KK'} = \frac{4}{3} \delta_{I(L} C_{MPQ)}.
\ee
This condition ensures that the matrix $Q_{IJ}$ is invertible, with inverse
\be
 Q^{IJ} = 2 X^I X^J - 6 C^{IJK} X_K,
\ee
where
\be
 C^{IJK} \equiv C_{IJK}.
\ee
We then have
\be
 X^I = \frac{9}{2} C^{IJK} X_J X_K.
\ee
The scalar potential is
\be
 {\cal V} = 27 C^{IJK} V_I V_J X_K
\ee
where $V_I$ are a set of constants. Without loss of generality we will assume $X^I,V_I>0$ and $C_{IJK} \geq 0$, so $\mathcal{V}>0$. It was shown in \cite{GR2} that the unique maximally supersymmetric solution of this theory is $AdS_5$ with vanishing vectors and constant scalars given by $X_I = \bar{X}_I \equiv V_I / \xi$, where $\xi^{3} = \frac{9}{2}C^{IJK}V_I V_J V_K$, with the radius of $AdS_5$ given by $g^{-1} \equiv (\xi \chi)^{-1}$. The above theory can be consistently truncated to minimal gauged supergravity as follows: $A^I=\bar{X}^{I} \mathcal{A}$ and $X^I=\bar{X}^I$.

In this paper,  we are interested in a particular $U(1)^3$ gauged supergravity that arises upon compactification of Type IIB supergravity on $S^{5}$. In the above language, this theory has $n=3$, $C_{IJK} = 1$ if $(IJK)$ is a permutation of $(123)$ and $C_{IJK} = 0$ otherwise, and $\bar{X}^I=1$ (so $\bar{X}_I=1/3$). In this case the action reduces to (\ref{u1cubed}).

\paragraph{Supersymmetric solutions.} The general nature of supersymmetric solutions of this theory was
deduced in \cite{GR2} following closely the corresponding analysis
for the minimal theory given in \cite{GG}. We will briefly summarise some of the results of this analysis. Given a supercovariantly
constant spinor $\epsilon$, one can construct a real scalar $f \sim
\bar{\epsilon} \epsilon$ and a real vector $V^\mu \sim
\bar{\epsilon} \gamma^\mu \epsilon$ and three real two forms $J^i_{\mu\nu} \sim \bar{\epsilon} \gamma_{\mu\nu} \epsilon$ where $i=1,2,3$. These obey $V^2 = -f^2$, so $V$
is timelike or null, and it turns out that $V$ is always Killing. Also note:
\begin{equation}\label{dJ}
dJ^{i} = 3\chi \epsilon^{1ij}V_{I}\left(A^{I} \wedge J^{j} + X^{I}\star_{5} J^{j} \right)
\end{equation}
which we will need later.
There are two cases: a ``null" case, in which $V$ is globally null
and a ``timelike" case in which $V$ is timelike in some region
${\cal U}$ of spacetime. The former case was treated in \cite{GG,GS}
and does not concern us here because such solutions cannot describe
black holes.

In the timelike case, we can, without loss of generality, assume that $f>0$ in ${\cal U}$, and introduce local coordinates so that the metric takes the form
\be
\label{eqn:metric}
 ds^2 = -f^2 \left( dt + \omega \right)^2 + f^{-1} h_{mn} dx^m dx^n,
\ee with $V = \partial/\partial t$, $h_{mn}$ is a metric on a
4-dimensional Riemannian ``base space" $\mathcal{B}$ and $\omega$ a 1-form on
$\mathcal{B}$. We choose the orientation on $\mathcal{B}$ so that $(dt + \omega) \wedge \eta_4$ is positively oriented in space-time, where $\eta_4$ is the volume form on $\mathcal{B}$. Fierz identities imply the $J^i$ are anti-self dual two-forms defined on $\mathcal{B}$ that satisfy the algebra of the unit quaternions -- hence $\mathcal{B}$ admits an almost hyper-K\"ahler structure. Supersymmetry then implies that the base space is K\"ahler with K\"ahler form $J^{1}$. Necessary and sufficient conditions for the existence of a supercovariantly constant spinor, in the timelike class, are derived in~\cite{GR2} and all take the form of equations defined on $\mathcal{B}$. We will not record all those conditions here, however we note that supersymmetry implies~\cite{GR2}
\be
\label{eqn:maxwell}
 F^I = d \left[ X^I f (dt + \omega) \right] + \Theta^I - 9 \chi f^{-1} C^{IJK} V_J X_K J^{1},
\ee
and
\be
\label{eqn:thetaconstr}
 X_I \Theta^I = - \frac{2}{3} G^+,
\ee
where $\Theta^I$ are self-dual 2-forms on $\mathcal{B}$ and
\be
 G^{\pm} = \frac{1}{2} f \left( d\omega \pm \star_4 d\omega \right),
\ee
where $\star_4$ is the Hodge dual on $\mathcal{B}$. The field equations of the theory are all satisfied once we impose the equations of motion for the Maxwell fields \cite{GR2}, i.e., the Bianchi identities
\be
 dF^I = 0,
\ee
and the Maxwell equations,
\be
 d \left( Q_{IJ} \star F^J \right) = - \frac{1}{4} C_{IJK} F^J \wedge F^K.
\ee
One can substitute the expression (\ref{eqn:maxwell}) into the Maxwell equation to obtain an equation on $\mathcal{B}$:
\begin{eqnarray}\label{maxonbase}
d \star_{4} d(f^{-1}X_{I}) &=& -\frac{1}{6}C_{IJK}\Theta^J \wedge \Theta^K + 2\chi V_{I} f^{-1} G^{-} \wedge J^{1} \nonumber \\ & +& 6\chi^2 f^{-2}(Q_{IJ}C^{JMN}V_{M}V_{N} + V_{I}X^{J}V_{J})\eta_{4}.
\end{eqnarray}

\subsection{Near-horizon Geometries}
We are interested in classifying the near horizon geometries of
supersymmetric black hole solutions of the above theory. The
strategy is to introduce coordinates adapted to the presence of a
Killing horizon, and then examine the conditions imposed by
supersymmetry and the field equations in the near horizon limit,
which we make precise below.   \par Following the reasoning
of~\cite{HDBH,GR1}, the line element of a supersymmetric black hole
may be written in \emph{Gaussian null coordinates} $(v,r,x^a)$:
\begin{equation}\label{gauss}
ds^2 = -r^2\Delta(r,x)^2dv^2 + 2dvdr + 2rh_{a}(r,x)dv dx^{a} + \gamma_{ab}(r,x)dx^{a}dx^{b},
\end{equation} with the horizon located at $r=0$. The supersymmetric Killing vector is $V = \partial / \partial v$ and for $r>0$ (the exterior of the black hole) is timelike, so $f=r\Delta$ and thus $\Delta>0$ for $r>0$ (although we will allow for $\Delta=0$ at $r=0$). Spatial cross sections of the horizon are given by $r=0$ and $v=$ constant: this defines a three-manifold, which we denote by $H$, with coordinates $x^a$. Since we are interested in black hole near-horizon geometries, ultimately we will require $H$ to be compact. The \emph{near-horizon limit} is defined by $r \rightarrow \epsilon r$ and $v \rightarrow v/\epsilon$ and $\epsilon \rightarrow 0$.  After this limit is taken, the functions $\Delta,h_a, \gamma_{ab}$ in the line element~(\ref{gauss}) depend only on $x^{a}$ (the coordinates on $H$).  As in~\cite{GR1}, first we will proceed by evaluating all equations as a power series in $r$ and take the near-horizon limit at the end of this section. \par \noindent Let the volume form $\eta_{3}$ of $\gamma_{ab}$ be chosen so that the spacetime volume form $\eta = dv \wedge dr \wedge \eta_{3}$ has positive orientation. Following~\cite{GR2} we work in a gauge where $i_V A^{I} = fX^{I}$.  We can then write
\begin{equation}\label{eqn:gaugepot}
A^I =  r\Delta X^{I}dv + A^{I}_r dr + a^{I}_{a}dx^{a}.
\end{equation} Note that $A^{I}_{r}$ does not survive the near-horizon limit.  The two forms $J^{i}$ may be written as~\cite{HDBH}
\begin{equation}\label{Ji}
J^{i} = dr  \wedge Z^{i} + r(h \wedge Z^{i} - \Delta \star_{3} Z^{i})
\end{equation} where $\star_{3}$ is the Hodge dual with respect to $\gamma_{ab}$ and the one-forms $Z^{i} = Z^{i}_{a}dx^{a} $ satisfy $\star_{3}Z^{i} = \frac{1}{2}\epsilon_{ijk}Z^{j} \wedge Z^{k}$ , i.e. they are orthonormal with respect to $\gamma_{ab}$.  Substituting~(\ref{Ji}) into~(\ref{dJ}) yields
\begin{eqnarray}
\label{dZeq}
\hat{d} Z^{i} &=& h \wedge Z^{i} -\Delta \star_{3}Z^{i} + r \partial_{r}(h \wedge Z^{i} - \Delta \star_{3}Z^{i}) \nonumber \\ &+& 3\chi\epsilon_{1ij}V_{I}\left[X^{I}\star_{3}Z^{j} + a^{I} \wedge Z^{j} - rA^{I}_r(h \wedge Z^{j} - \Delta \star_{3} Z^{j}) \right]
\end{eqnarray} and
\begin{eqnarray}
\star_{3} \hat{d}h & = & \Delta h + \hat{d}\Delta + r\star_{3}(h
\wedge
\partial_{r} h) - r(\partial_{r}\Delta) h + 2r\Delta \partial_{r}h +
r\Delta^2 \epsilon_{ijk}Z^{i} \langle Z^{j},\partial_{r}Z^{k}\rangle
\nonumber \\ &+&  6\Delta \chi V_{I} \left( X^{I} +r \Delta
A_{r}^{I} \right)Z^{1}.
\end{eqnarray} where $\hat{d}$ is the exterior derivative on $H$~\cite{GR1}. These equations closely resemble the minimal gauged case~\cite{GR1}, the main difference being the presence of the scalar fields.  To leading order in $r$ the expressions become:
\begin{equation}
\label{dZ}
\hat{d}Z^{i} = h \wedge Z^{i} - \Delta\star_{3}Z^{i} + 3\chi\epsilon_{1ij}V_{I}\left[X^{I}\star_{3}Z^{j} + a^{I}\wedge Z^{j} \right] + \mathcal{O}(r)
\end{equation} and
\begin{equation}
\label{dh}
\star_{3}\hat{d}h = \Delta h + \hat{d}\Delta + 6\chi\Delta V_{I}X^{I}Z^{1} + \mathcal{O}(r).
\end{equation}  For $r>0$, $\omega$ can be defined by $i_V \omega=0$ and $d\omega = -d(f^{-2}V)$ so in this coordinate system,
\begin{equation}
\omega = -\frac{dr}{\Delta^2r^2} - \frac{h}{\Delta^2r}.
\end{equation}  We next compute\footnote{Note that $i_{V}\star_{5}\lambda_{p} = f^{p-1}\star_{4}\lambda_{p}$ for a $p$-form on the base manifold}
\begin{equation}
G^{+} = \frac{1}{2}\left(f d\omega + i_V\star_{5}d\omega \right),
\end{equation} which gives
\begin{equation}\label{gplus}
G^{+} = dr \wedge \mathcal{G} + r(h \wedge \mathcal{G} + \Delta
\star_{3} \mathcal{G})
\end{equation} where
\begin{eqnarray}
\mathcal{G}
& = & -\frac{3\hat{d}\Delta}{2r\Delta^2} +
\frac{3(\partial_{r}\Delta) h}{2\Delta^2} - \frac{3
\partial_{r}h}{2\Delta} - \frac{1}{2} \epsilon_{ijk}Z^{i}\langle
Z^{j},\partial_{r}Z^{k}\rangle - \frac{3\chi
V_{I}}{r\Delta}\left[X^{I} + \Delta r A_{r}^{I}\right]Z^{1}.
\end{eqnarray} We now turn to the determination of the Maxwell fields~(\ref{eqn:maxwell}).
Following~\cite{Gut} write
\begin{equation}
\label{Theta}
\Theta^{I} = -\frac{2}{3}X^{I}G^{+} + N^{I}
\end{equation} where $X_{I}N^{I} = 0$ and since $N^{I}$ is self dual on
the base it can be written as
\begin{equation}
N^{I} = dr \wedge T^{I} + r(h\wedge T^{I} + \Delta \star_{3}T^{I})
\end{equation}
for some $T^I=T^I_adx^a$.
Substituting into~(\ref{eqn:maxwell}), we find
\begin{eqnarray}
F^{I} &=& \left[\partial_{r}(r\Delta X^{I})dr + r\hat{d}(\Delta
X^{I}) \right]\wedge dv + dr \wedge Q^{I} + r h\wedge Q^{I} \nonumber \\
      &+& r\Delta\star_{3}T^{I} - X^{I}\star_{3}h
      -rX^{I}\star_3 \partial_{r}h
      -\frac{2}{3}rX^{I}\Delta\star_{3}Z^{i}\epsilon_{ijk}\langle
      Z^{j},\partial_{r}Z^{k}\rangle \nonumber \\
      &+& \chi V_{J} \star_{3}Z^{1} \left[9C^{IJK}X_{K} -
      4X^{I}\left(X^{J} + r\Delta A_{r}^{J}\right)\right]
\end{eqnarray} where
\begin{eqnarray}
Q^{I} &=& \frac{\hat{d}X^{I}}{r\Delta} - \frac{\partial_{r}X^{I}
h}{\Delta} + \frac{1}{3}X^{I}Z^{i}\epsilon_{ijk}\langle
Z^{j},\partial_{r}Z^{k} \rangle + T^{I} \nonumber \\ &+&
\frac{1}{r\Delta}\chi V_{J}\left[2X^{I}(X^{J} + r\Delta
A_{r}^{J})-9C^{IJK}X_{K}\right]Z^{1}.
\end{eqnarray}  Note that since we
require $F^{I}$ to be regular on the horizon, then so must $Q^{I}$
be\footnote{In minimal gauged supergravity it was found that the Maxwell field was automatically regular on the horizon~\cite{GR1}. With extra vector multiplets though, it seems we need this as an extra (reasonable) assumption -- this was also found in the ungauged theory~\cite{Gut}. Note, however, $X_IF^I$ is automatically regular.}. Reading off the
the $x^a$ components of $F^{I}$ we get:
\begin{eqnarray}
\label{F}
\hat{d}a^I &=& rh \wedge Q^{I} + r\Delta
\star_{3}Q^{I} - \star_{3}\hat{d}X^{I} + r\partial_{r}X^{I}
\star_{3}h - r\Delta X^{I}\epsilon_{ijk}\star_{3}Z^{i}\langle
Z^{j},\partial_{r}Z^{k} \rangle \nonumber \\ &-& X^{I} \star_{3}h
- rX^{I}\star_{3}\partial_{r}h  + 6\chi V_{J}\left[3C^{IJK}X_{K} -
X^{I}(X^{J} + r\Delta A_{r}^{J})\right] \star_{3}Z^{1} \nonumber \\
&=& -\star_3 \hat{d}X^I- X^I \star_3h +6\chi V_J( 3C^{IJK}X_K-X^IX^J)\star_3Z^1 +\mathcal{O}(r)
\end{eqnarray}
where the last equality follows from regularity of $Q^I$ at $r=0$. The Bianchi identity contracted with $X_I$ implies
\begin{equation}\label{divh}
\hat{d}\star_{3}h + \frac{2}{3}Q_{IJ}\hat{d}X^{I}\wedge \star_{3}\hat{d}X^{J} + 2\chi V_{K}X^{K} \hat{d}\star_{3}Z^1 + 4\chi V_{K}\hat{d}X^{K} \wedge \star_{3} Z^{1} = \mathcal{O}(r).
\end{equation} In the ungauged case, $\chi =0 $ and therefore integrating~(\ref{divh}) over ${H}$ leads to $\hat{d}X^{I} = 0$, since $Q_{IJ}$ is a positive definite metric on the scalar manifold~\cite{Gut}. However, in the gauged theory this conclusion cannot be drawn and indeed we will find explicit solutions where the scalars are \emph{not} constant on ${H}$.  \par  We next turn to the computation of the spin connection of $\gamma_{ab}$ using~(\ref{dZeq}), which allows us to deduce
\be\label{covZ}
\nabla_a Z^i_b = -\frac{\Delta}{2}( \star_3 Z^i)_{ab}+ \gamma_{ab}( h \cdot Z^i+ 3\chi V_IX^I \delta^1_i)- Z^i_a h_b -3\chi V_IX^IZ^i_aZ^1_b+3\chi V_Ia^I_a \epsilon_{1ij}Z^j_b + \mathcal{O}(r).
\ee
where $\nabla$ is the connection of $\gamma_{ab}$. From~(\ref{covZ}) it is easy to deduce that
\be
\label{divZ}
\star_3d \star_3 Z^i= \nabla_b Z^{ib} = 2 h \cdot Z^i+6\chi V_IX^I\delta^i_1+3 \chi V_I\epsilon_{1ij}a^I \cdot Z^j + \mathcal{O}(r).
\ee Finally, we can determine the Ricci tensor $R_{ab}$ of $\gamma_{ab}$,  using
\begin{equation}
R_{ab}Z^{ib} = \nabla_b \nabla_a Z^{ib} - \nabla_a\nabla_b Z^{ib}.
\end{equation} After an involved calculation, we find:
\begin{eqnarray}
\label{ricci}
R_{ab} &=& \left(\frac{\Delta^2}{2} + h^2 + 4\chi V_{J}X^{J}h \cdot Z^1 + \frac{2}{3}Q_{IJ}\hat{d} X^{I}\cdot \hat{d} X^{J} + 4\chi V_{J}Z^{1c}\partial_{c}X^{I} + 2\chi^2[6(V_{I}X^{I})^2 - \mathcal{V}] \right)\gamma_{ab} \nonumber \\
&-&\nabla_{(a}h_{b)} - h_{a}h_{b} - 6\chi V_{I}X^{I}h_{(a}Z^{1}_{b)} - 6\chi V_{I} \partial X^{I}_{(a}Z^{1}_{b)} - 2\chi^2 Z^{1}_{a}Z^{1}_{b}\left(9(V_{I}X^{I})^2 - \mathcal{V}\right) + \mathcal{O}(r).
\end{eqnarray} Note that all gauge dependent terms cancel, as they must.
Our final task is to examine the Maxwell equation~(\ref{maxonbase}). We find that as in the minimal theory this imposes no new constraints. More precisely we find that the Maxwell equation leads to a second order equation for the scalars which can also be derived by taking the $\hat{d}$-derivative of (\ref{F}) and using (\ref{divh}) and (\ref{divZ}). There are no further conditions imposed by supersymmetry or the field equations. \par  We are interested in determining near-horizon solutions. Accordingly, as discussed in the beginning of this section, we shall henceforth work strictly in the near-horizon limit. This amounts to dropping all $\mathcal{O}(r)$ terms in the equations derived above and setting $r=0$ in all other quantities. Therefore all equations are now defined purely on $H$ and from now one we will denote the exterior derivative on $H$ simply $d$.

\subsection{Some general results}
We have not been able to solve the near-horizon equations derived in the previous section in full generality. Indeed, it is not even known how to solve the near-horizon equations in full generality in minimal gauged supergravity~\cite{GR1,KLR2}. However, much like in the minimal case we have obtained some general results which allow one to classify near-horizon geometries into two classes: static ($V \wedge dV \equiv 0$) or non-static. The following lemmas provide conditions for this: \\

\noindent {\it Lemma 1}. The following conditions are equivalent: (a) $V \wedge dV \equiv 0$, (b) $dh=0$, (c) $\Delta \equiv 0$.

\noindent {\it Proof}. Assume (a). The $rab$ components of $V \wedge
dV \equiv 0$ give $dh=0$ so (a) implies (b). Now assume (b). If
$\Delta$ is nonzero then equation (\ref{dh}) can be solved for $Z^1$
which leads to $Z^1 \wedge dZ^1
=0$. Then equation (\ref{dZ}) implies $Z^1 \wedge \star_3 Z^1=0$ which contradicts $Z^1$ having unit norm. Therefore (b) implies (c). Finally assume (c). Equation (\ref{dh}) implies $dh \equiv 0$. But $\Delta \equiv 0$ and $dh \equiv 0$ implies $V \wedge dV \equiv 0$. Therefore (c) implies (a). \\

\noindent {\it Lemma 2}. If $\Delta$ vanishes at a point then $\Delta$ vanishes everywhere.

\noindent {\it Proof}. Taking the divergence of (\ref{dh}) and using
(\ref{divh}) and (\ref{covZ}) gives: \bea \nabla^2 \Delta &=& -
\left( h+ 6 \chi V_IX^I Z^1 \right) \cdot \nabla \Delta \nonumber \\
&+& \left[ \frac{2}{3}Q_{IJ} \partial X^I \cdot \partial X^J -2\chi
V_I \partial X^I \cdot Z^1 - 8\chi V_IX^I( h \cdot Z^1 +3 \chi
V_IX^I) \right]\Delta \eea
and therefore one sees that this equation for $\Delta$ is of the same form as in the minimal case. This allows one to repeat the argument used in~\cite{KLR2} to prove that if $\Delta$ vanishes at $p$ then all derivatives of $\Delta$ at $p$ also vanish. Hence by analycity we deduce that $\Delta \equiv 0$. \\
We should point out that lemma 2 will not be used to derive any of the results in this paper and thus in particular we will not need to assume analycity on the horizon.

\subsubsection{A special case}
\label{sec:special}
There is a special case in which the near-horizon equations can be solved, without the assumption of rotational symmetries. This case is specified by: $h$ Killing, $\Delta = \textrm{constant}$ (possibly zero), $X^I= \textrm{constant}$ and $h=-3\chi V_IX^I Z^1$. It is not obvious that these conditions are consistent, however it turns out they are. These conditions will arise later in our analysis of $U(1)^2$-invariant near-horizon solutions and are the analogues of the assumptions made in the analysis for minimal gauged supergravity~\cite{GR1}. The steps turn out to parallel the minimal case~\cite{GR1} very closely. Define $W= Z^2+iZ^3$. In general, with no assumptions, equation (\ref{dZ}) implies:
\be
\label{W}
dW = \left[ h+3\chi V_ IX^I -i\Delta Z^1-3\chi iV_I a^I \right] \wedge W
\ee
and thus $W \wedge dW=0$. Therefore locally we can write $W= \sqrt{2} F dw$ for some complex functions $F$ and $w$ on ${H}$.  Although $w$ is gauge invariant, $F$ is not and we choose to work in a gauge where $F$ is real. As in~\cite{GR1}, we can define real coordinates $(x,y,z)$ by $w= (x+iy)/\sqrt{2}$ and $Z^1 = \partial/\partial z$. The metric in these coordinates reads:
\be
ds_3^2 = (dz+\alpha)^2+ 2F^2 dwd\bar{w}
\ee
where $\alpha= \alpha_w dw + \alpha_{\bar{w}} d \bar{w}$ is a real one-form.
So far we have not used any of our assumptions. From the assumptions it follows that $Z^1$ is Killing and thus $\alpha$ and $F$ are independent of $z$. From the equation for $dZ^1$ (\ref{dZ}) it follows that:
\be
\partial_w \alpha_{\bar{w}} - \partial_{\bar{w}} \alpha_w = -i\Delta F^2
\ee
just like in the minimal case. Thus we have determined $\alpha$ in terms of $F$ (up to a gradient which can be absorbed into the definition of $z$). Equation (\ref{W}) can be solved for $V_Ia^I$ to get:
\be
V_I a^I=-\frac{\Delta}{3\chi} {Z^1}- \frac{i}{3\chi}\left( \frac{\partial_{w} F}{F}dw -\frac{\partial_{\bar{w}} F}{F}d\bar{w} \right)
\ee
and substituting this into (\ref{F}) gives:
\be
\partial_w \partial_{\bar{w}} \log F^2 =  \left( 2\chi^2 \mathcal{V}- 9 \chi^2 (V_IX^I)^2 - \Delta^2 \right) F^2
\ee which is Liouville's equation. There are three cases to consider
depending on whether the RHS is positive, zero or negative.
Define $g^2\lambda \equiv -2\chi^2 \mathcal{V}+ 9 \chi^2
(V_IX^I)^2$. Importantly, $\lambda$ can be positive, negative or
vanish, depending on the values of the scalars. Note that in minimal supergravity $\lambda=-3$; we will see that it is possible to get
some qualitatively different geometries by taking $\lambda \geq 0$
which have no counterpart in the minimal case. First let us deduce
some conditions for which $\lambda \geq 0$. For simplicity work in
$U(1)^3$ supergravity in which case\footnote{In the QFT literature this is known as the K\"allen function.}: \be \lambda(X^1,X^2,X^3) = -2(X^1X^2 +X^2X^3
+X^1X^3) + (X^1)^2 +(X^2)^2+(X^3)^2. \ee It has the property that if
$\sqrt{X^1}+\sqrt{X^2} \leq \sqrt{X^3}$ (or permutations of $(123)$)
then $\lambda \geq 0$.

One can repeat the steps of~\cite{GR1} to get:
\begin{enumerate}
\item $\Delta^2=-g^2\lambda$ which is only possible for scalars satisfying $\lambda \leq 0$. In this case Liouville's equation reduces to the wave equation. By a holomorphic change of coordinates it is then possible to set $F=1$ and in these coordinates $\alpha= \frac{g \sqrt{-\lambda}}{2}( ydx-xdy)$, so the horizon becomes: \be ds_3^2= \left( dz + \frac{g\sqrt{-\lambda}}{2}(ydx-xdy) \right)^2 +
dx^2+dy^2 \ee which for $\lambda <0$ is the homogeneous metric on
Nil and generalises the solution found in the minimal
case~\cite{GR2}. However we can also take $\lambda=0$ now, in which
case the horizon is locally isometric to $R^3$. This solution
has no counterpart in minimal gauged supergravity.

\item  $\Delta^2< -g^2\lambda$ which is only possible for $\lambda<0$. Liouville's equation can be solved, and then a holomorphic change of coordinates leads to $F^2 = -\frac{1}{(g^2\lambda+\Delta^2)x^2}$ and $\alpha = -\frac{\Delta dy}{(g^2\lambda+\Delta^2)x}$. The geometry is the homogeneous metric on $SL(2,R)$:
\be ds_3^2= \left( dz + \frac{\Delta}{|\Delta^2+ g^2 \lambda | )}
\frac{dy}{x} \right)^2 + \frac{1}{| \Delta^2 +g^2 \lambda  |} \left(
\frac{dx^2+dy^2}{x^2} \right). \ee Note that $\Delta=0$ is allowed
in this case: this corresponds to the horizon being locally
isometric to $R \times H^2$.
\item $\Delta^2> -g^2\lambda$ which is possible for any $\lambda$. After a homomorphic change of coordinates we have:
\be F^2 = \frac{2}{(\Delta^2+g^2\lambda) (1+w \bar{w})^2}, \qquad
\alpha = -\frac{i\Delta}{(\Delta^2+g^2\lambda)(1+w \bar{w})} \left(
\frac{dw}{w}-\frac{d\bar{w}}{\bar{w}} \right). \ee Now introduce
real coordinates via $w=\tan( \theta /2 )e^{i\psi}$. For $\Delta>0$
one can write: \be \label{selfdual} ds_3^2 = \frac{1}{\Delta^2
+g^2\lambda} \left[ \frac{\Delta^2}{\Delta^2+g^2\lambda}( d\phi
+\cos\theta d\psi)^2 + d\theta^2 +\sin^2\theta d\psi^2 \right] \ee
where $z= \Delta \phi/(\Delta^2+g^2\lambda)$. This is the
homogeneous geometry of a squashed $S^3$. The full near-horizon
geometry is that of the black hole solutions found
in~\cite{GR2}\footnote{The near-horizon limit of their solution is
parameterized by three independent constants called $q_I$, which are
related to our parameters by $q_I= 4X_I/(\Delta^2+g^2\lambda)$.}.
One can also have $\Delta=0$ when $\lambda > 0$; this gives: \be
ds_3^2 = dz^2 + \frac{1}{g^2 \lambda} ( d\theta^2 + \sin^2\theta
d\phi^2 ) \ee which is locally isometric to $R \times S^2$.  This
solution has no counterpart in minimal supergravity.
\end{enumerate}
This analysis has been purely local. We are mainly interested in a compact horizon $H$. It is easy to see that the $\lambda=0$ case of 1 above gives a $T^3$ horizon, whereas the $\Delta=0$ case of 3 gives an $S^1 \times S^2$ horizon.

\section{$U(1)^2$-invariant near-horizon geometries}
Consider an asymptotically $AdS_5$ black hole admitting two
rotational symmetries $m_1$ and $m_2$. The near-horizon solution
will inherit these symmetries. Therefore we are interested in
classifying all near-horizon solutions for which there are two
commuting Killing fields $m_1$ and $m_2$ on $H$ that preserve
$h,\Delta$, the Maxwell fields $F^I$ and the scalars $X^I$.

Under these conditions a welcome simplification occurs for the
Maxwell fields. A standard argument, which uses the Bianchi identity
for $F^I$ as well as the fact that the Lie derivatives of $F^I$
along the Killing fields vanish, tells us that
$F^I_{\mu\nu}m_1^{\mu}m_2^{\nu}$ is a constant. Since we are looking
for solutions which are asymptotically globally $AdS_5$, both
Killing fields vanish at a (different) point. Therefore
$F^I_{\mu\nu} m_1^{\mu} m_2^{\nu}=0$. This condition is inherited in
the near-horizon limit.

We can choose local coordinates $x^a=(\rho,x^i)$ with $\partial / \partial x^i$ Killing, so that the metric on $H$ is:
\be
\gamma_{ab}= d\rho^2 +\gamma_{ij}(\rho) dx^idx^j
\ee
and $\Delta$, $X^I$ and the components of $h$ and $F^I$ are functions only of $\rho$. We will allow $\partial / \partial x^i$ to be arbitrary linear combinations of $m_i$ and thus they need not have closed orbits. We will enforce the fact that $m_i$ have closed orbits once we have determined the local form of the solution.

We will define a positive function $\Gamma(\rho)$ and a one-form
$k_i(\rho)dx^i$ by: \be \label{h} h=\Gamma^{-1}k_i dx^i -
\frac{\Gamma'}{\Gamma} d\rho \ee where a prime denotes a derivative
with respect to $\rho$. The components of the Maxwell fields on $H$
can be written as: \be \label{Bdef} \frac{1}{2}{F}^I_{ab} dx^a
\wedge dx^b = B^I_i(\rho) d\rho \wedge dx^i \ee where we have used
the fact that $F^I_{ij}=0$ argued above.

Taking the dual of equation (\ref{F}) and using (\ref{Bdef}) leads to:
\be
\label{Feq}\label{eqn:B}
 \star_2 B^I = (X^I)' d\rho+ X^Ih -6\chi V_J( 3C^{IJK}X_K-X^IX^J) Z^1
\ee
where $\star_2$ denotes the Hodge star with respect to the two-dimensional metric $\gamma_{ij}$ (with volume form $\eta_2$ such that $\eta_3=d\rho \wedge \eta_2$). Contracting with $X_I$ implies
\be
\label{Z}
Z^1 = \frac{1}{2\chi V_IX^I}( \star_2B - h)
\ee
where for convenience we have defined $B \equiv X_IB^I$. Now we can read off the $\rho$ and $i$ components of (\ref{Feq}); the $\rho$ component leads to
\be
\label{XIup}
 {X^I}'+
\frac{2\Gamma'}{\Gamma} X^I- \frac{9 C^{IJK} V_JX_K }{V_LX^L}
\frac{\Gamma'}{\Gamma}=0
\ee
whereas the $i$ component leads to an expression for $B^I$ in terms of $B$. Contracting with $V_I$ gives:
\be
\label{VX} (V_IX^I)'+ \frac{2\Gamma' V_IX^I}{\Gamma}- \frac{\Gamma'
\mathcal{V}}{3\Gamma V_IX^I}=0. \ee

Using (\ref{Z}) and (\ref{VX}), the $\rho i$ component of equation
(\ref{ricci}) simplifies considerably leaving: \be 0=R_{\rho i}= -
\frac{1}{2} \Gamma^{-1}\gamma_{ij}(k^j)' \ee and hence \be
\label{kvec} k^i \equiv \gamma^{ij}k_i= \textrm{constant}. \ee

Next, using (\ref{Z}), equation (\ref{dh}) gives:
\bea
\label{Deltaprime}
\Delta'+ \frac{2\Delta \Gamma'}{\Gamma} &=& 0, \\
(\Gamma^{-1}k)' +2 \Delta \star_2 (\Gamma^{-1}k) &=& -3\Delta B
\label{B} \eea  where for a one form $\omega_i(\rho)dx^i$ we defined $\omega'=
\omega_i'(\rho) dx^i$. Hence \be \Delta =
\frac{\Delta_0}{\Gamma^2} \ee
where $\Delta_0$ is a non-negative constant.
Substituting (\ref{Z}) into (\ref{dZ}) leads to: \bea \label{dZij} \Gamma^{-1} k \wedge \star_2 B-
\frac{\Delta \Gamma'}{\Gamma} \star_2 1 &=& 0, \\
\label{dZij2} \left[ \frac{1}{V_IX^I}( \star_2 B- \Gamma^{-1}k)
\right]' &=& \frac{\Delta}{ V_IX^I}( B+ \star_2 \Gamma^{-1}k) -
\frac{1}{V_IX^I} \frac{\Gamma'}{\Gamma} \star_2 B.\eea

Now let us turn to equation (\ref{covZ}). The
$\rho\rho$ component gives: \be \label{rhorho} h \cdot Z^1 + 3\chi
V_IX^I = \frac{1}{2\chi V_IX^I} \left[ \frac{\Gamma''}{\Gamma}-
\frac{\Gamma'^2}{2\Gamma^2} - \frac{\Gamma'}{\Gamma}
\frac{(V_IX^I)'}{V_IX^I} \right] \ee and the $ij$ component gives: \be
\label{ij} \frac{ \Gamma'}{4\chi V_IX^I \Gamma} \gamma_{ij}' =
-\frac{\Delta \Gamma'}{4\chi V_IX^I \Gamma} \sqrt{\gamma}
\epsilon_{ij} + \gamma_{ij}( h\cdot Z^1+ 3\chi V_IX^I )- Z^1_i( h+3\chi
V_IX^IZ^1)_j. \ee

There are two qualitatively distinct cases depending on whether $\Gamma$ is a constant or not.

\paragraph{Constant $\Gamma$ case:}  From equations (\ref{h}) and (\ref{kvec}) one can see that $h$ must be Killing, equation (\ref{Deltaprime}) implies $\Delta = \textrm{constant}$ and equation (\ref{XIup}) implies $X^I = \textrm{constant}$. Further equations (\ref{rhorho}) and (\ref{ij}) imply $h=-3\chi V_IX^I Z^1$. This turns out to be a very similar case to the one studied in the minimal case~\cite{GR1} which can be solved without the assumption of the $U(1)^2$ symmetry. This is the special case we solved in section \ref{sec:special}.

\paragraph{ Non-constant $\Gamma$ case:} In the minimal theory we found that $\Gamma$ was a more convenient coordinate than $\rho$ on $H$~\cite{KLR2}. Due to the presence of the scalar fields and scalar potential we will see that a better coordinate emerges, although $\Gamma$ will still be useful. We first address solving the scalar equation (\ref{XIup}). From this one can deduce an equation for $X_I$:
\be \label{XI} X_I'+ \frac{\Gamma'}{\Gamma}X_I -
\frac{\Gamma'}{\Gamma}\frac{V_I}{V_LX^L}=0 \ee and therefore an
equation for the scalar potential: \be \mathcal{V}'+
\frac{\Gamma'}{\Gamma} \mathcal{V}- \frac{6 \xi^3\Gamma'}{\Gamma
V_LX^L}=0. \ee Now define the positive function $x(\rho)$ by: \be x
\equiv \frac{\Gamma \mathcal{V}}{6\xi^2}. \ee The equation for the
scalar potential can now be written as: \be \label{pot} x'=
\frac{\xi\Gamma'}{V_IX^I} \ee We will find that the function $x$
turns out to be a better coordinate that $\Gamma$ (observe that
$x=\Gamma$ in the minimal limit). Indeed equation (\ref{VX}) may be
written as: \be \frac{d}{dx}(\Gamma^2V_IX^I)= 2\xi x \ee which
integrates to: \be \label{VXx} \Gamma^2V_IX^I = \xi x^2 +\xi c_1 \ee
where $c_1$ is some integration constant.  Further, plugging
(\ref{pot}) into (\ref{VXx}) leads to a differential equation
relating the two coordinates $x$ and $\Gamma$ which is easily
integrated to: \be \Gamma^3 = H(x) \equiv  x^3 + 3c_1 x + c_2 \ee
where $c_2$ is an integration constant and thus from (\ref{VXx}) we
have determined $V_IX^I$ as a function of $x$. Equation (\ref{XI})
can be written as: \be \frac{d}{dx}(H^{1/3} X_I)= \frac{V_I}{\xi}
\ee which then integrates to: \be  X_I = H(x)^{-1/3}\left( \frac{V_I
x}{\xi } +K_I \right)\ee where $K_I$ are integration constants. If
one calculates $\mathcal{V}$ from this expression for $X_I$ one
learns that $C^{IJK}V_IV_JK_K=0$. Further the constraint relating
the scalars then tells us that: \be c_1 =
\frac{9}{2\xi}C^{IJK}V_IK_JK_K, \qquad c_2 =
\frac{9}{2}C^{IJK}K_IK_JK_K. \ee Hence we have now fully determined
the scalars in terms of $x$.

Since we are now assuming $\Gamma$ (and hence $x$) is not a constant, we can derive a number of useful results from equations (\ref{rhorho}) and (\ref{ij}). Firstly (\ref{rhorho}) gives:
\be \label{kZ} (k \cdot Z^1) = \frac{1}{2\chi V_IX^I}\left[
\Gamma'' + \frac{\Gamma'^2}{2\Gamma}-\frac{\Gamma' (V_IX^I)'}{\Gamma
V_IX^I} \right] - 3\chi V_I X^I\Gamma  = \frac{\xi \chi}{\Gamma} \frac{dy}{dx}\ee where the last equality follows from (\ref{pot}) and we have defined \be y=
\frac{1}{4\xi^2\chi^2}\Gamma x'^2 - \Gamma^3. \ee
Multiply (\ref{ij}) by $\gamma^{ij}$ and use (\ref{rhorho}) to get:
\be
\frac{\Gamma''}{\Gamma} = \frac{\Gamma'}{2\Gamma} (\log \gamma)' +
\frac{\Gamma'}{\Gamma} \frac{(V_IX^I)'}{V_IX^I} \ee which integrates
to: \be \label{detgamma} \sqrt{\det \gamma_{ij}} = \beta^2 |x'| \ee using (\ref{pot}) and where $\beta$ is a positive constant
(chosen to match with the minimal limit~\cite{KLR2}). \\

In order to make further progress it is now necessary to split the analysis into two cases: $\Delta_0>0$ and $\Delta_0=0$.

\subsection{Non-static near-horizon geometry}
From lemma 1 we see that this case corresponds to $\Delta_0>0$. We also see that the constants $k^i$ cannot both vanish; if they did, then $dh=0$ and hence by lemma 1 $\Delta=0$, contradicting $\Delta_0>0$. Assume that $\Gamma$ (and thus $x$) is not constant -- we have already dealt with the $\Gamma$ constant case.

Eliminating $B$ between equations (\ref{B}) and (\ref{dZij}) leads
to: \be k^i(\Gamma^{-1}k_i)' = \frac{3 \Delta_0^2 \Gamma'}{\Gamma^4}
\ee and since $k^i$ are constants one can integrate this to get: \be
\label{normk} k^ik_i = C^2\Gamma - \frac{\Delta_0^2}{\Gamma^2} \ee
where $C$ is a positive constant. Now, contract (\ref{ij}) with
$k^ik^j$ and use (\ref{normk}) to get: \be \label{kZeq} (k \cdot
Z^1)^2 = C^2\Gamma - \frac{\Delta_0^2}{\Gamma^2} - \frac{C^2
x'^2}{4\xi^2\chi^2 \Gamma}. \ee Now, eliminate $k \cdot Z^1$ between
equations (\ref{kZ}) and (\ref{kZeq}) to get: \be \left(
\frac{dy}{dx} \right)^2 + \frac{C^2}{\xi^2\chi^2}y =
-\frac{\Delta_0^2}{\xi^2\chi^2} \ee which integrates to \be
y=-\frac{C^2}{4 \xi^2\chi^2}\left(x- \alpha_0 \right)^2 -
\frac{\Delta_0^2}{C^2}. \ee where $\alpha_0$ is an integration
constant. This implies: \be \label{xprime1} x'^2 = \frac{4\xi^2
\chi^2 P(x)}{\Gamma} \ee where \be \label{P1} P(x) = H(x) -
\frac{C^2}{4\xi^2 \chi^2}\left( x-\alpha_0 \right)^2 -
\frac{\Delta_0^2}{C^2}. \ee

Eliminate $B$ between equations (\ref{B}) and (\ref{Z}) to get: \be
\label{ZDeltapos} Z^1= \frac{1}{2\chi V_IX^I }\left[ -\frac{1}{3
\Delta} \star_2 (\Gamma^{-1}k)' -\frac{1}{3}\Gamma^{-1} k +
\frac{\Gamma'}{\Gamma} d\rho \right]. \ee Now we will use the
$GL(2,R)$ freedom associated with the $x^i$ coordinates to set
$k^1=1$ and $k^2=0$. Note that equation (\ref{normk}) implies: \be
\label{gamma11} \gamma_{11} = C^2\Gamma-
\frac{\Delta_0^2}{\Gamma^2}. \ee Plugging our expression for $y$
back into (\ref{kZ}) and then equating this to $k^i$ times the $i$
component of (\ref{ZDeltapos}) gives the following ODE: \be
\frac{d}{d\Gamma} \left( \frac{\gamma_{12}}{\gamma_{11}} \right) =
\frac{\Delta_0 \beta^2}{[ C^2\Gamma^3-\Delta_0^2]^2} \left(
\frac{\mathcal{V}}{6\xi^2}( 2C^2\Gamma^3+\Delta_0^2) -
3C^2\alpha_0\Gamma^2 - (C^2\Gamma^3-\Delta_0^2)
\frac{\Gamma}{6\xi^2} \frac{d \mathcal{V}}{d\Gamma} \right) \ee
where we have used (\ref{pot}) and (\ref{detgamma}). Thankfully this
integrates in a similar way to the minimal case to give: \be
\frac{\gamma_{12}}{\gamma_{11}} = \frac{\Delta_0 \beta^2 \left(
\alpha_0 - x \right)}{C^2\Gamma^3-\Delta_0^2} \ee plus some
integration constant which we may set to zero using some of the
remaining $GL(2,R)$ freedom. We may use the remaining freedom to set
$\beta=1$. Now using (\ref{detgamma}) allows us to deduce
$\gamma_{22}$. Therefore we have completely determined the 2-metric
$\gamma_{ij}$ in terms of $x$. It is thus convenient to use $x$,
rather than $\rho$ as the 3rd coordinate on $H$; $\gamma_{xx}$ can
be deduced from (\ref{xprime1}). The full near-horizon geometry in
the non-static case is now determined.  The final step is to
determine the field strengths. From~(\ref{eqn:B}) and
(\ref{ZDeltapos}) we can deduce the components of $F^I$ on $H$:
\begin{equation}
\frac{1}{2}F^I_{ab} dx^a \wedge dx^b =
d\left[H^{-1/3}X^{I}\left(\Delta_{0}dx^{1} +
(x-\alpha_{0})dx^{2}\right) \right]
\end{equation}  The field strengths $F^{I}$ are then fully determined from the gauge potentials~(\ref{eqn:gaugepot}) upon taking the near-horizon limit.  We now summarise our results for the non-static case below:

\paragraph{Summary of non-static near-horizon solutions:}
\begin{enumerate}[(i)]
\item  If $\Gamma$ (and hence $x$) are not constant:
\bea
\label{horx}
\gamma_{ab}dx^adx^b &=& \frac{H(x)^{1/3} dx^2}{4\xi^2 \chi^2 P(x)} +
H(x)^{-2/3}\left( C^2H(x)-\Delta_0^2 \right) \left( dx^1 + \frac{
\Delta_0 \left(
\alpha_0 -x \right)}{C^2H(x) -\Delta_0^2}dx^2 \right)^2 \nonumber \\
&+& \frac{ 4\chi^2\xi^2 H(x)^{1/3} P(x)}{C^2 H(x)-\Delta_0^2} (dx^2)^2, \\
\Delta &=& \frac{\Delta_0}{H(x)^{2/3}}, \qquad k= \frac{\partial}{\partial x^1}, \qquad h= H(x)^{-1/3}k - \frac{H'(x)}{3H(x)} dx \\ \label{scalar}
X_I &=& H(x)^{-1/3} \left( \frac{V_I}{\xi} x +K_I \right) \\
A^{I}  &=&  \frac{X^I}{H^{1/3}} \left[\Delta_{0}\left(H^{-1/3}rdv +
dx^{1}\right) + (x-\alpha_0)dx^{2}\right] \eea where
$H(x)=x^3+3c_1x+c_2$ and $P(x)= H(x) - \frac{C^2}{4\xi^2
\chi^2}\left( x-\alpha_0 \right)^2 - \frac{\Delta_0^2}{C^2}$ and
$C,\Delta_0$ are positive constants, $\alpha_0$ an arbitrary
constant and $c_1=\frac{9}{2\xi}C^{IJK}V_IK_IK_K$,
$c_2=\frac{9}{2}C^{IJK}K_IK_JK_K$ where $K_I$ are constants
satisfying $C^{IJK}V_IV_JK_K=0$.

\item If $\Gamma$ is a constant, the scalars are a constant, $\Delta$ is a constant and the metric on $H$ must be one of: the homogeneous metrics on the group manifold $Nil$, $SL(2,R)$ or squashed $S^3$ depending on the value of $\Delta$, see section \ref{sec:special}. The latter case arises as the near-horizon limit of the asymptotically $AdS$ black hole solutions found in~\cite{GR2}, as explained
in section \ref{sec:special}.
\end{enumerate}

The near-horizon geometry of the supersymmetric black holes
of~\cite{KLR} is non-static with non-constant $\Gamma$ and hence
must be described by (i). We will prove this below.

\subsection{Static near-horizon geometry}
Lemma 1 tells us this corresponds to $\Delta_0=0$. We will now analyse the
$\Gamma$ not constant case, as we dealt with the $\Gamma$ constant case earlier. It is now possible to have $k^i=0$. This is dealt with
in the appendix.

Thus, now assume that $k^i$ are not both zero. Equation (\ref{B})
implies $\Gamma^{-1}k_i$ are constants. Since $k^i$ are constants we
can define a positive constant $C$ by $C^2= \Gamma^{-1}k_ik^i$. Use
the $GL(2,R)$ freedom to set $k^1=1$ and $k^2=0$. Therefore
$\gamma_{11}$ and $\gamma_{12}$ are both a constant times $\Gamma$
and hence we can use some of the remaining $GL(2,R)$ freedom to set
$\gamma_{12}=0$ and therefore $\gamma_{11}= k^ik_i=C^2\Gamma$. We
can now repeat some of the steps used in the non-static case.
Namely, contract (\ref{ij}) with $k^ik^j$ and use (\ref{normk}) to
get: \be \label{kZeq2} (k \cdot Z^1)^2 = C^2\Gamma - \frac{C^2
x'^2}{4\xi^2\chi^2 \Gamma}. \ee Now, eliminate $k \cdot Z^1$ between
equations (\ref{kZ}) and (\ref{kZeq2}) to get: \be \left(
\frac{dy}{dx} \right)^2 + \frac{C^2}{\xi^2\chi^2}y = 0 \ee which
integrates\footnote{This equation has another solution: $y=0$. However, it can be shown this implies $k=0$, which is the case we consider in the appendix.} to \be y=-\frac{C^2}{4 \xi^2\chi^2}\left(x- \alpha_0
\right)^2. \ee where $\alpha_0$ is an integration constant. This
implies: \be \label{xprime2} x'^2 = \frac{4\xi^2 \chi^2
P(x)}{\Gamma} \ee where \be \label{P2} P(x) = H(x) -
\frac{C^2}{4\xi^2 \chi^2}\left( x-\alpha_0 \right)^2. \ee Observe
that all these equations can be obtained from the $\Delta_0 \to 0$
limit of the corresponding equations in the $\Delta_0>0$ case. We
can now use (\ref{detgamma}) to deduce $\gamma_{22}$ and hence have
fully determined the near-horizon geometry in this case. It remains
to deduce the Maxwell fields. Observe that equation (\ref{dZij})
implies that $\star_2 B \propto k$ and thus $Z^1_i \propto k_i$.
Using the solution for $y$ and substituting into (\ref{kZ}) implies
that: \be Z^1_i= -\frac{(x-\alpha_0)k_i}{2\chi\xi \Gamma^2} \ee and
substituting this into (\ref{eqn:B}) leads to the components of the
Maxwell field on $H$: \be \frac{1}{2}F^I_{ab}dx^a \wedge dx^b= d[
H(x)^{-1/3}X^I(x-\alpha_0) dx^2 ]. \ee Thus we see that the
components of the field strengths on $H$ can also be obtained as the
$\Delta_0 \to 0$ limit of the non-static case. We now summarise the
static-near horizon solutions:

\paragraph{Summary of static near-horizon solutions:}


\begin{enumerate}[(i)]
\item If $\Gamma$ is a constant, the scalars are constants. $H$ is either locally isometric to $R \times H^2$ ($\lambda<0$), $R^3$ ($\lambda=0$) or $R \times S^2$ ($\lambda >0$); see section (\ref{sec:special}). The first is the near-horizon limit of a supersymmetric black ``string''~\cite{Klemm}.
\item  $\Gamma$ (and hence $x$) are not constant and $k^i$ not both zero. This leads to a solution which can be obtained by taking the $\Delta_0 \to 0$ limit of (\ref{horx}), which amounts to simply setting $\Delta_0=0$ in the solution (\ref{horx}).
\item $\Gamma$ not constant,  $k^i=0$. This case is analysed in the Appendix, where the local form of the solution is given. It contains a case with zero Maxwell fields and non-constant scalars. In the case of zero Maxwell fields and constant scalars, $H$ is locally isometric to $H^3$ and the near-horizon solution locally to $AdS_5$.
\end{enumerate}

As discussed in section \ref{sec:special} and below, the $R \times S^2$ example in solution (i) above may be
compactified to describe the near horizon of a regular
supersymmetric black ring. We shall see that the solution (ii) also
describes the near-horizon geometry of a supersymmetric black ring
but suffers from a conical singularity.

\subsection{Global analysis}\label{sec:global}
The preceding analysis has been entirely local. We are primarily
interested in those solutions that arise from the near-horizon limit
of black holes, and hence we must restrict attention to solutions
for which $H$ is compact.  This turns out to be a strong constraint,
as a compact three-dimensional manifold with $U(1) \times U(1)$
isometry must be homeomorphic to $T^3, S^1 \times S^2,  S^3$, or a
lens space~\cite{gowdy}.

Consider first the static near horizon solutions.  The $R \times
H^2$ possibility in (i) is immediately excluded since $H^2$ cannot
be compactified without breaking the rotational symmetries.
Furthermore, the subcase in (iii) which has zero Maxwell fields and constant scalars, has $H$ locally isometric to $H^3$, and hence cannot be compactified without
breaking the rotational symmetries. For the non-static geometries,
(ii) is ruled out apart from the case where $H$ is isometric to a
homogeneously squashed $S^3$. As we have already explained in
section \ref{sec:special}, this solution is the same as the
near-horizon limit of the black holes of~\cite{GR2}.
\par  Consider now the geometries that are locally $R^3$ or $R
\times S^2$ in case (i) of the static solutions. It is clear that we
may compactify these to yield compact horizons with $T^3$ and $S^1
\times S^2$  geometry respectively. We emphasize that these cases
only occur for particular (constant) values of the scalars, and
\emph{cannot} exist in minimal gauged supergravity.  Further there
are no known black hole solutions with such horizon geometries. \par
Finally, consider the remaining possibilities, all of which have $x$
non-constant.  We first note that because $\partial/\partial x^1$ is
a linear combination of $m_1,m_2$, its norm,  $\gamma_{11}$, is a
scalar invariant.  By definition $x > 0$ and further $dx/d\Gamma >
0$. Hence for all cases, $\gamma_{11}$ is a monotonically increasing
function of $x$.  Therefore $x$ is uniquely determined in terms of
$\gamma_{11}$ and is a globally defined function on $H$. Compactness
of $H$ then implies that $x$ must achieve a distinct minima and
maxima on $H$. Hence the one-form $dx$ must vanish at two different
positive values of $x$.  Computing $(dx)^2$ for the near horizons
with non-constant $x$, we find this excludes case (iii) (see Appendix), leaving (i) and (ii) of the non-static and static solutions
respectively as the only possibilities. We conclude any solution with compact $H$ and
non-constant $x$ must be given by~(\ref{horx}), whether non static
($\Delta_0 > 0$) or static ($\Delta_0 = 0$).
\par For these solutions,
\begin{equation}
(dx)^2 = \frac{4\chi^2\xi^2 P(x)}{\Gamma},
\end{equation} so we must impose that $P(x)$ be non-negative and have at least two distinct positive roots $x_1,x_2$ with $P(x) > 0$ for $x_1 < x < x_2$.  It is straightforward to see that this implies $P(x)$ must have another distinct positive root $x_3$ such that $x_1 < x_2 < x_3$ (compactness excludes $x_2 = x_3$).  These conditions constrain the parameters of the solution. For example, the positivity of the scalars $X_I$ (\ref{scalar}) then tell us that $K_I > -V_Ix_1/\xi$. Furthermore, note that $x$ is defined only up to a multiplicative constant.  Hence,  one of the parameters of~(\ref{horx})  may be removed by a suitable rescaling of $x$.  Explicitly, for some constant $\Omega > 0$, (\ref{horx}) is invariant under
\begin{equation}
\label{scale}
x \rightarrow \Omega x, \qquad x^{1} \rightarrow  \Omega^{-1} x^1, \qquad  K_{I} \rightarrow \Omega K_I, \qquad C^2 \rightarrow \Omega C^2, \qquad \Delta_{0} \rightarrow \Omega^2 \Delta_{0}, \qquad \alpha_{0} \rightarrow \Omega\alpha_0
\end{equation}  We now turn to a detailed analysis of these two cases.

\subsubsection{$\Delta_0=0$: Unbalanced black ring}
From~(\ref{horx}) we can read off the metric on $H$:
\bea
\label{horring}
\gamma_{ab}dx^adx^b &=& \frac{H(x)^{1/3} dx^2}{4\xi^2 \chi^2 P(x)} +
C^2H(x)^{1/3} (dx^1)^2
+ \frac{4\xi^2\chi^2  P(x)}{C^2 H(x)^{2/3}} (dx^2)^2.
\eea It is clear we must remove the conical singularities at $x = x_1$ and $x=x_2$ at which the Killing field $\partial / \partial x^2$ vanishes. If this were possible, by suitably fixing the period of $x^2$, then the metric~(\ref{horring}) would describe a regular horizon with topology $S^1 \times S^2$, with $S^1$ and $S^2$  parameterised by $x^1$ and $(x, x^2)$ respectively, i.e. the horizon of  a supersymmetric black ring in $AdS_{5}$.   \par
The condition for removing the conical singularities at $x_1,x_2$ in the compact 2-manifold covered by $(x,x^2)$ is
\begin{equation}\label{singular}
\frac{H(x_2)}{H(x_1)}  = \left(\frac{x_3-x_2}{x_3-x_1} \right)^2.
\end{equation} The RHS is obviously less than unity. But since
 $H'(x)>0$ (which follows from the fact $d\Gamma / dx>0$), $H(x)$ is a monotonically increasing function of $x$ and the LHS of (\ref{singular}) is larger than one. So, although $H$ has $S^1 \times S^2$ topology, it necessarily has a conical singularity at one of the poles of the $S^2$. If~(\ref{horring}) represents the near horizon of a black ring, then that black ring would be unbalanced, i.e. require external forces to prevent it from self-collapse.   \par Let us consider the full five-dimensional geometry further. If we define a new radial coordinate $R = H(x)^{-1/3}r$, the spacetime metric is given by
\begin{equation}
\label{ads3}
ds^2 = H(x)^{1/3}\left[-C^2R^2dv^2 + 2dvdR + C^2\left(dx^1 + Rdv\right)^2 \right]+ \frac{H(x)^{1/3}dx^2}{4\xi^2 \chi^2 P(x)} + \frac{4\xi^2\chi^2  P(x)}{ C^2H(x)^{2/3}} (dx^2)^2
\end{equation}
The part of the metric in the square brackets is locally isometric
to $AdS_{3}$.  Therefore the full five dimensional metric is a
warped product of $AdS_{3}$ with a squashed $S^2$ with a singularity
at one of its poles. Note that in the limit of vanishing gauge coupling $\chi \rightarrow 0$ this solution reduces to $AdS_{3} \times S^2$, the near-horizon geometry of an asymptotically flat supersymmetric black ring~\cite{multibpsring}. This near-horizon geometry can be oxidised on
$S^5$ to 10d, where it can be made regular with a different
topology. The horizon topology becomes $S^1 \times M_7$, where $M_7$ is some complicated compact manifold (see Appendix). The special case $K_I=0$ (which is a solution to the
minimal theory) lifts to the solution of \cite{ads3}, which was
shown to lead to a discrete set of regular warped $AdS_{3}$
geometries. In the Appendix we outline how to do this for the more
general case. However, these regular solutions cannot be reduced to
5d and therefore do not have an interpretation in terms of 5d black
holes. Indeed, an interesting question is whether there exist 10d asymptotically $AdS_5 \times S^5$ black holes whose near-horizon geometry is given by these regular warped $AdS_3$ geometries.

\subsubsection{$\Delta_0>0$: Topologically spherical black hole}
The horizon metric is
\begin{equation}\label{nonstatic}
 \gamma_{ab}dx^a dx^b = \frac{H(x)^{1/3} dx^2}{4\xi^2 \chi^2 P(x)} + H(x)^{-2/3}\left[A(x)(dx^1 + \omega(x)dx^2)^2 + B(x)(dx^2)^2 \right]
\end{equation} where
\begin{equation}
A(x) = C^2\left( P(x) + \frac{C^2}{4\chi^2\xi^2}(x-\alpha_0)^2 \right) \qquad B(x) = \frac{4\chi^2\xi^2H(x)P(x)}{A(x)} \qquad \omega(x) = \frac{\Delta_{0}(\alpha_0 -x)}{A(x)}
\end{equation} Recall $x_1 \leq x \leq x_2$. Note that $\gamma_{11} > 0$ unless $\alpha_0 = x_i$ where $i=1,2$. We will consider $\alpha_0 \neq x_i$ now and deal with the degenerate cases $\alpha_0=x_i$ in the Appendix -- it turns out they can be obtained as the $\alpha_0 \rightarrow x_i$ limits of the generic case.
The 2-metric $\gamma_{ij}$ induced on surfaces of constant $x$ is non-degenerate everywhere except at the endpoints $x=x_i$, where the Killing vectors $\omega(x_i)\partial_{x_1} - \partial_{x_2}$ vanish.  Using $P(x_i) = 0$, we find
\begin{equation}
\omega(x_i) = \frac{4\chi^2\xi^2}{C^4(\alpha_0 - x_i)}
\end{equation} and thus $\omega(x_1) \neq \omega(x_2)$. This implies that the Killing field which vanishes at $x=x_1$ is distinct from the one which vanishes at $x=x_2$. To avoid conical singularities these Killing fields must generate rotational symmetries, i.e. have closed orbits.  Accordingly, they must be proportional to the $m_i$ and we write
\begin{equation}
m_i = -d_i\left(\omega(x_i)\frac{\partial}{\partial x^1} - \frac{\partial}{\partial x^2} \right)
\end{equation}
for some constants $d_i$. Define coordinates $\phi_i$ such that $m_i=\partial / \partial \phi_i$ and $\phi_i \sim \phi_i +2\pi$. The coordinate change from $(x^1,x^2)$ to $(\phi_1,\phi_2)$ is given by
\be\label{coordchange}
x^i = -[ \omega(x_1)d_1 \phi_1 + \omega(x_2)d_2 \phi_2], \qquad x^2 = d_1\phi_1+d_2 \phi_2.
\ee Absence of conical singularities then fixes the constants $d_i$ up to a sign:
\begin{equation}
d_{i}^2 = \frac{A(x_{i})}{4\chi^4\xi^4 P'(x_i)^2}.
\end{equation} Note that using $P(x_i)=0$ one can show that
\begin{equation}
A(x_{i}) = \frac{C^4(x_i - \alpha_0)^2}{4\chi^2\xi^2}.
\end{equation}  The solution is now globally regular: $H$ has $S^3$ topology with $m_{1}$ vanishing at $x_1$ and $m_2$ vanishing at $x_{2}$. In the appendix we show that the coordinate change from $(x^1,x^2)$ to $(\phi_1,\phi_2)$ is also valid in the special cases $\alpha_0=x_i$ (although $d_{i} \to 0$ as $\alpha_0 \to x_i$, $d_{i}\omega(x_i)$ is non-zero in this limit). Therefore we need not treat this case separately anymore.

\paragraph{Relation to known black hole.}
 Next we show that the near-horizon limit of the black holes of~\cite{KLR} are in fact isometric to the non-static solution we have derived in this paper~(\ref{nonstatic}). Observe that one can deduce the near-horizon limit of the black hole in~\cite{KLR} from section 2.7 of that paper. The near-horizon geometry we derived is parameterised by $(C^2,\alpha_0,\Delta_0,K_I)$ with one constraint $C^{IJK}V_JV_KK_I=0$. We can rewrite this solution in terms of the parameters $(x_1,x_2,x_3,K_I)$ which are related by
\bea
\label{Calpha}
C^2 &=& 4g^2(x_1+x_2+x_3), \qquad \alpha_0= \frac{x_1x_2+x_1x_3+x_2x_3 -3c_1}{2(x_1+x_2+x_3)}, \\ \label{Delta0}
\Delta_0^2 &=& C^2(x_1x_2x_3+c_2) -\frac{C^4}{4g^2}\alpha_0^2.
\eea Note that the roots $x_i$ are not arbitrary positive real numbers such that $x_1<x_2<x_3$: they are further constrained by $\Delta_0^2>0$. Now, the scale transformation (\ref{scale}) can be used to ensure that $g(x_3-x_1)/\Delta_0 <1$. Use the remaining scale transformations on $x$ (i.e. ones with $\Omega >1$) to fix:
\be
\label{xfix}
\frac{g}{\Delta_0}( x_1+x_2+x_3) = \left( 1+\sqrt{1-\frac{g}{\Delta_0}(x_3-x_2)} + \sqrt{1-\frac{g}{\Delta_0}(x_3-x_1)} \right)^2
\ee
where the factors under the square roots are positive as a consequence of the ordering of the roots and $g(x_3-x_1)/\Delta_0<1$. This allows one to define two positive constants $a,b$ by:
\be
\label{abdef}
\Xi_a \equiv 1-a^2g^2= \frac{g(x_3-x_2)}{\Delta_0}, \qquad \Xi_b \equiv 1-b^2g^2= \frac{g(x_3-x_1)}{\Delta_0}
\ee
so $0<b<a<g^{-1}$.
The equations (\ref{xfix}) and (\ref{abdef}) can be inverted:
\be
\label{xofab}
x_1 = g\Delta_0
\left( b^2+ \frac{2r_m^2}{3} \right), \qquad x_2= g \Delta_0 \left(
a^2+ \frac{2r_m^2}{3} \right), \qquad x_3 = \frac{\Delta_0}{g} \left(1+
\frac{2g^2r_m^2}{3}\right) \ee where $r_m^2 \equiv g^{-1}(a+b) + ab>0$. Also define constant $e_I$:
\be
\label{edef}
e_I = \frac{2r_m^2}{3} +
\frac{K_I}{g\Delta_0},
\end{equation}
and observe that the constraint on the $K_I$ translates to $3 \bar{X}^Ie_I=2r_m^2$ (as in~\cite{KLR}). Equation (\ref{edef}) implies:
\be
\label{cofbeta}
c_1 = \frac{g^2 \Delta_0^2}{3}\left( \beta_2 - \frac{4r_m^4}{3} \right), \qquad c_2 = g^3\Delta_0^3 \left( \beta_3 - \frac{2r_m^2 \beta_2}{3} + \frac{16 r_m^6}{27} \right)
\ee
where $\beta_2 = \frac{27}{2}C^{IJK} \bar{X}_Ie_J e_K$ and $\beta_3= \frac{9}{2}C^{IJK}e_Ie_Je_K$. Substituting equations (\ref{xofab}) and (\ref{cofbeta}) into (\ref{Calpha}) and (\ref{Delta0}) allows one to solve for $\Delta_0$:
\begin{equation}
\Delta_{0} = \frac{1}{2g^2 \Xi_a \Xi_b \sqrt{\delta}}
\end{equation}
where
\bea
\nonumber \delta &=& \frac{1}{4g^2 \Xi_a^2 \Xi_b^2}[4g^2 (1+ag+bg)^2 \beta_3 - g^4 \beta_2^2 + 2g^2 (g^2 a^2b^2 + a^2 +b^2) \beta_2 \\ &-& g^2(g^2 a^2 b^2-2gab^2 -2ga^2b + (a-b)^2 ) r_m^4]
\eea
is the same $\delta$ appearing in \cite{KLR}; note $\delta>0$ is equivalent to $\Delta_0^2>0$. Thus we have determined $\Delta_0$ in terms of $a,b,e_I$. This proves (looking at (\ref{xofab})) that the parameters $x_1,x_2,x_3,K_I$ can be written uniquely in terms of $a,b,e_I$ (this was not obvious as we defined $a,b,e_I$ as functions of $x_1,x_2,x_3,K_I$).
Now define a new coordinate $\theta$ by
\be
\label{thetax}
\sin^2\theta = \frac{x_2-x}{x_2-x_1}
\ee so the range $0 \leq \theta \leq \pi/2$ covers the entire range of $x$. The endpoints $x=x_1,x_2$ correspond to $\theta = \pi/2,0$ respectively. This can be inverted:
 \begin{equation} x= x_1\sin^2\theta +x_2\cos^2\theta =g\Delta_0\left( \rho(\theta)^2 +
\frac{2r_m^2}{3}\right)  \end{equation}  where $\rho(\theta) = a^2\cos^2\theta + b^2\sin^2\theta$ and the second equality follows from (\ref{xofab}). This implies
\begin{equation}
P(x) = \Delta_0^3g(a^2-b^2)^2 \Delta_{\theta} \sin^2\theta \cos^2\theta, \qquad H(x) = g^3\Delta_0^3 F(\rho^2)
\end{equation} where $\Delta_{\theta} = 1 - g^2\rho(\theta)^2$ and $F(\rho^2) \equiv \rho^6 +2r_m^2 \rho^4+ \beta_2 \rho^2 +\beta_3$.
This leads to \be F(\rho^2)^{1/3} \frac{d\theta^2}{\Delta_{\theta}}
= \frac{H(x)^{1/3} dx^2}{4g^2 P(x)}. \ee Note that this verifies
that the $\theta\theta$ component of the near-horizon limit of the
black hole in~\cite{KLR} is equal to the $xx$ component of our
non-static near-horizon metric. For completeness, we give the
expressions for $d_i$ and $\omega(x_i)$:
\begin{eqnarray}
d_1 &=& \frac{2b^4g^2 + 4g^2b^2r_m^2 + g^2b^2a^2 + b^2 - a^2 + g^2\beta_2}{2g\Xi_b(a^2-b^2)}  \\
d_2 &=& -\frac{2a^4g^2 + 4g^2a^2r_m^2 + g^2b^2a^2 - b^2 + a^2 + g^2\beta_2}{2g\Xi_a(a^2-b^2)} \\
\omega(x_1) &=& -\frac{1}{2g\Delta_0^3(4g^2b^2r_m^2 + 2b^4g^2 + g^2b^2a^2 + b^2-a^2 + g^2\beta_2)(1+ag+bg)^2} \\
\omega(x_2) &=& -\frac{1}{2g\Delta_0^3(4g^2a^2r_m^2 + 2g^2a^4 + g^2b^2a^2 + a^2-b^2+g^2\beta_2)(1+ag+bg)^2}.
\end{eqnarray}
We have checked explicitly that our non-static near horizon solution (\ref{nonstatic}) written in terms of the coordinates $(\theta, \phi_1,\phi_2)$ and the parameters $(e_I, a, b)$ is exactly the same as the near horizon limit of the solutions~\cite{KLR} in their $(\theta,\psi,\phi)$ notation, provided we identify $\phi_1 = \psi$ and $\phi_2 = \phi$.

\section{Discussion}
In this work, we have derived the most general near horizon geometry
of a regular, supersymmetric asymptotically $AdS_{5}$ black hole
solution of $U(1)^3$ gauged supergravity with two rotational
symmetries. We emphasize that all known five dimensional black hole
solutions, including black rings, have two rotational symmetries. It
has been suggested that stationary black holes with only one
rotational symmetry could exist~\cite{HDBH}, essentially because the higher dimensional analogue of the rigidity
theorem only guarantees this~\cite{rigidity}. However, in this work we are interested in
supersymmetric black holes\footnote{The rigidity
theorem~\cite{rigidity} is only valid for non-extremal black
holes.}. In five dimensional ungauged supergravity the analysis is
much simpler and one can in fact solve the near-horizon equations
generally. It turns out that in that theory a supersymmetric
near-horizon geometry must possess two rotational symmetries after
all (in fact it must be a homogeneous space)~\cite{HDBH,Gut}.

The near-horizon geometries we derived all possess enhanced symmetry
$SO(2,1) \times U(1)^2$ and are 1/2 BPS (from the point of view of
minimal supergravity). The bosonic part of this is guaranteed on
general grounds from~\cite{KLR3} (in fact it follows from extremality
rather than supersymmetry). However the supersymmetry of these
backgrounds is also enhanced. We expect this to also hold on general
grounds, although no proof of this has been given. A recent
classification of 1/2 BPS geometries~\cite{half} revealed that only
one $U(1)$ spatial isometry is guaranteed. Therefore the validity of
our assumption of two rotational symmetries is still unclear\footnote{Note that solutions to gauged supergravity with only one spatial $U(1)$ symmetry have been found in~\cite{her}.}. It
would be nice to lift this assumption from our analysis.

Our results, summarized in Section 2, imply that the most general
near horizon geometry of a topologically \emph{spherical} black
hole, with two rotational symmetries, is given by the near horizon
geometry of the four-parameter black hole solutions of~\cite{KLR}. Therefore, any other topologically spherical
supersymmetric black hole must either have fewer than two
rotational symmetries, or have the same near horizon geometry (and
hence the same horizon geometry) as the solution of~\cite{KLR}. Note that as
the gauge coupling is turned off the black hole solution~\cite{KLR} reduces\footnote{To see this one must rescale the parameters of the black hole in an appropriate manner.} to the BMPV
black hole, which also depends on four parameters and has a
topologically spherical horizon.

In contrast
with the corresponding analysis in minimal gauged
supergravity~\cite{KLR2}, we have been unable to rule out the
existence of black holes with $T^3$ or $S^1 \times S^2$ horizons;
these must have near-horizon geometries $AdS_{3}\times T^2$ and $AdS_{3}\times
S^2$ respectively. These solutions, which have no analogue in the
minimal theory, have constant scalars and can only
occur in regions of this moduli space satisfying
$\lambda(X^1,X^2,X^3) \geq 0$ where $\lambda$ is given
in~(\ref{lambda}).

An interesting question is whether a supersymmetric $AdS_5$  black
ring actually exists in this theory. If there is such a solution,
and it possesses two rotational symmetries, we have shown that it
must have the near-horizon geometry $AdS_3 \times S^2$ (given by (\ref{NHring})), as discussed above. This solution is not generic in the sense that the constant scalars must take values such that $\lambda>0$, which in particular excludes any such solutions from having equal charges (i.e. occuring in minimal gauged supergravity). As
noted in Section 2, it is a three parameter solution. It can be parameterised by $(L, q^{I})$
where $L$ is the radius of the $S^1$ on the horizon and $q^I$ are
the dipole charges defined by
\begin{equation}
  q^{I} = \frac{1}{4\pi} \int_{S^2} F^{I}.
 \end{equation}
For our near-horizon solution (\ref{NHring}) it turns out that
\begin{equation}
 \sum_{I=1}^{3} q^{I} = -\frac{1}{g}.
\end{equation}
In contrast, the near horizon geometry of the analogous
supersymmetric black ring of ungauged $U(1)^3$ supergravity, which can also
be parameterised by an analogous set $(L,q^I)$, has unconstrained dipoles and therefore four independent parameters. The radius of the $S^2$ in (\ref{NHring}) is $(g\sqrt{\lambda})^{-1}$ and can be made small compared to the AdS radius $g^{-1}$ (as $\lambda$ can be large compared to $1$). However, as one turns off the gauge coupling (and hence sends the AdS radius to infinity) its size will diverge. This assumes that one does not rescale the scalars (and hence gauge fields) by factors of $g$; indeed, when taking such ``flat space'' limits, usually one only considers possible rescalings of the parameters and coordinates of the solution and not the fields (e.g. as for the spherical black hole discussed above).
Therefore, for these reasons, we conclude that if~(\ref{NHring}) is indeed the near horizon
of a supersymmetric black ring in $AdS_{5}$, it would not reduce (as
one turns off the gauge coupling $g$) to the analogous solution in
the ungauged theory. \par Moreover, as in the minimal case~\cite{KLR2}, we found a near
horizon geometry that is a warped product of $AdS_{3}$ with a
squashed $S^2$ that suffers from a conical singularity at one of its
poles. Unlike the above case, in the limit $g \rightarrow 0$, it \emph{does} reduce to $AdS_{3}\times
S^2$, the near horizon of an asymptotically flat black ring. Therefore it seems natural to consider this as the generalisation of the near-horizon geometry of the black ring of ungauged supergravity, despite it being singular. This
four-parameter solution could arise as the near horizon of a
supersymmetric  \emph{unbalanced} black ring. If so this would mean that
rotation and electromagnetic forces are not sufficient to counteract
the gravitational self-attraction while simultaneously satisfying
the BPS equality. However, this suggests that by increasing the
angular momenta in such a way as to break the BPS condition, one
might be able to balance this configuration and construct a
non-supersymmetric black ring.
\\

\noindent {\bf Acknowledgments}. HKK and JL are supported by PPARC.
We would like to thank Mukund Rangamani and Harvey Reall for comments on the draft.

\appendix


\section{$\Delta_{0}=0$ and $k=0$}
Here we deal with a special case $\Delta_0=0=k^i$. Equation (\ref{rhorho}) becomes:
\be
\label{00case}
\frac{\Gamma''}{\Gamma}+ \frac{\Gamma'^2}{2\Gamma^2} - \frac{ \Gamma' (V_IX^I)'}{\Gamma V_IX^I} = 6\chi^2 (V_IX^I)^2
\ee
and therefore we see that $\Gamma$ cannot be a constant. Therefore we may use $\Gamma$ as a coordinate and write (\ref{00case}) as:
\begin{equation}
\frac{d}{d\Gamma} \left(\frac{\Gamma \Gamma'^2}{4\chi^2(V_I X^I)^2} - \Gamma^3 \right) = 0
\end{equation} and integrate to get
\begin{equation}
\Gamma'^2 = \frac{4\chi^2 (V_I X^I)^2 Q(\Gamma)}{\Gamma}
\end{equation} where $Q(\Gamma) = \Gamma^3 - \Gamma_0^3$ and $\Gamma_0$ is a real integration constant. Integrating~(\ref{dZij2}) we find
\begin{equation}
\star_2 B = \frac{V_I X^I \omega}{\xi \Gamma}
\end{equation} where $\omega_i$ are constants (possibly zero).
Since $Z^1$ has unit norm, we must have
\begin{equation}
\omega_i \omega^i = \frac{4\chi^2\xi^2 \Gamma_0^3}{\Gamma}
\end{equation}
and thus $\omega_i=0$ if and only if $\Gamma_0=0$.
Now, equation (\ref{ij}) simplifies to \be
\label{ijsc}\frac{\Gamma'}{4\chi V_IX^I \Gamma} \gamma_{ij}' = \chi
V_IX^I \left( 1+ \frac{2\Gamma_0^3}{\Gamma^3} \right) \gamma_{ij} -
\frac{3 V_IX^I}{4\chi \xi^2 \Gamma^2} \omega_i\omega_j. \ee Now we can perform a $GL(2,R)$ transformation on the coordinates $x^i$ to set $\omega_2=0$ (this is of course trivial if $\omega_i=0$). This equation then simplifies to: \be
\frac{ d}{d\Gamma} \log \gamma_{i2} = \frac{Q'(\Gamma)}{Q(\Gamma)} -
\frac{2}{\Gamma} \ee and thus $\gamma_{12}$ and $\gamma_{22}$ are
the same function up to a multiplicative constant. Hence we can use
more of the $GL(2,R)$ freedom to set $\gamma_{12}=0$ and solve for
$\gamma_{22}= Q(\Gamma)/\Gamma^2$. Now multiplying (\ref{ijsc}) by
$\gamma^{ij}$ gives: \be \frac{d}{d\Gamma} \log \gamma =
\frac{Q'(\Gamma)}{Q(\Gamma)} - \frac{1}{\Gamma} \ee and hence
$\gamma = C^2 Q(\Gamma)/\Gamma$ for some constant $C>0$. Thus we have determined the local form of the metric; converting to the $x$ coordinate introduced in the main text gives: \be \label{kzero}ds^2_3 = \frac{ H(x)^{1/3} dx^2}{4\chi^2 \xi^2
Q(x)} + C^2H(x)^{1/3} (dx^1)^2 +
\frac{Q(x)}{ H(x)^{2/3}} (dx^2)^2 \ee where $Q(x)= H(x)
-\Gamma_0^3$.

\par As discussed in the main text, for a compact horizon the one-form $dx$ must vanish at two distinct points (if $x$ is non-constant which is the case here). However, we have
\begin{equation}
(dx)^2 = \frac{4\chi^2\xi^2 Q(x)}{H^{1/3}}
\end{equation} and hence $dx$ can vanish at most at one point (for $\Gamma_0>0$ this point is defined by $\Gamma=\Gamma_0$; for $\Gamma_0=0$ there is no such point). Therefore~(\ref{kzero}) cannot correspond to a compact horizon. If $K_I = 0$ (in which case the scalars are constants) and $\omega_i=0$ (in which case the Maxwell fields vanish) the horizon metric is locally isometric to hyperbolic space $H^3$, and the full near horizon geometry is then locally $AdS_5$.

\section{The cases $\alpha_0 = x_i$}
Here we perform a global analysis of the cases $\Delta_0>0$ and $\alpha_0 = x_i$.  Consider the horizon metric~(\ref{nonstatic}). For clarity first consider $\alpha_0=x_1$. Note that in this case $A(x)$ vanishes at $x_1$ and is positive otherwise, and $B(x)$ vanishes only at $x_2$ and is positive otherwise. It follows that the 2-metric $\gamma_{ij}$ induced on surfaces of constant $x$ is non-degenerate everywhere on the interval $x_1 \leq x \leq x_2$ except at $x=x_1$, where the Killing field $\partial /\partial x^1$ vanishes, and at $x=x_2$, where the Killing field $\omega(x_2)\partial /\partial x^1 - \partial /\partial x^2$ vanishes. Hence these Killing fields describe rotational symmetries and must have closed orbits. Accordingly they must be proportional to the $m_i$ and we may write
\begin{equation}
m_1 = c_1 \frac{\partial}{\partial x^1}, \qquad m_2 = -d_2\left(\omega(x_1)\frac{\partial}{\partial x^1} - \frac{\partial}{\partial x^2} \right)
\end{equation} where $c_1, d_2$ chosen so that $m_{i} = \partial/\partial \phi_i$ and $ \phi \sim \phi + 2\pi$. Removing the conical singularity at $x=x_1$ we find
\begin{equation}
c_1^2 = \frac{\Delta_0^2}{\chi^2\xi^2 C^4 H'(x_1)^2}.
\end{equation} Removing the conical singularity at $x=x_2$ gives
\begin{equation}
d_2^2 = \frac{C^4(x_2-x_1)^2}{16\chi^6\xi^6 P'(x_2)^2}
\end{equation} The horizon metric is now globally regular and has $S^3$ topology with $m_1$ vanishing at the pole $x=x_1$ and $m_2$ vanishing at the other pole $x=x_2$. The coordinate change $(x^1,x^2) \rightarrow (\phi_1,\phi_2)$, given by
\begin{equation}
x^1 = c_1\phi_1 - d_2\omega(x_2)\phi_2 \qquad x^2 = d_2\phi_2,
\end{equation} may be obtained from~(\ref{coordchange}) by taking the limit $\alpha_0 \rightarrow x_1$.

Note that the $\alpha_0=x_2$ case is analogous and can also be made globally regular resulting in $S^3$ topology. One finds $m_i$ vanishes at $x=x_i$ where now:
\be
m_1=-d_1\left(\omega(x_1)\frac{\partial}{\partial x^1} - \frac{\partial}{\partial x^2} \right), \qquad m_2 =c_2 \frac{\partial}{\partial x^1}
\ee
and
\be
d_1^2 = \frac{C^4(x_2-x_1)^2}{16\chi^6\xi^6P'(x_1)^2}, \qquad c_2^2= \frac{\Delta_0^2}{\chi^2\xi^2 C^4 H'(x_2)^2}
\ee
and
\be
x^1=-d_1\omega(x_1)+\phi_1 c_2\phi_2, \qquad x^2= c_2\phi_2
\ee
which can be obtained from the $\alpha_0 \to x_1$ limit of (\ref{coordchange}).

\section{10d near-horizon geometries}
The 5d near-horizon geometries we have derived in $U(1)^3$ gauged
supergravity can all be lifted to solution of type IIB supergravity
using~\cite{Cvetic}\footnote{Note that the $X_i$ of~\cite{Cvetic}
are equal to our $X^I$; the field strengths of~\cite{Cvetic} $F^i$
are the same as ours $F^I$. Indeed the action (\ref{u1cubed}) agrees
with the one they give in~\cite{Cvetic} with the field
identifications just described.}: \bea ds_{10}^2 &=& W^{1/2} ds_5^2
+ W^{-1/2} \sum_{I=1}^3 (X^I)^{-1} [d\mu_I^2 + \mu_I^2 (d\varphi_I
+g A^I)^2 ], \\ \label{fiveform}
F_5 &=& (1+\star_{10})\left( 2g \sum_{I=1}^3 ((X^I)^2 \mu_I^2 -W X^I) \epsilon_5 - \frac{1}{2g} \sum_{I=1}^3 (X^I)^{-1} \star_5 dX^I \wedge d\mu_I^2 \right. \nonumber \\
&{}& \left. \qquad \quad \qquad + \frac{1}{2g^2} \sum_{I=1}^3
(X^I)^{-2} d\mu_I^2 \wedge (d\phi_I +gA^I) \wedge \star_5 F^I
\right), \eea where $W= \sum_{I=1}^3 \mu_I^2 X^I>0$, and $\mu_I,
\varphi_I$ are coordinates on $S^5$ such that $\sum_{I=1}^3
\mu_I^2=1$.

It can occur that certain singular 5d solutions can actually be made
regular when oxidised to 10d. This is indeed the case for the static
near-horizon solution which we found (\ref{ads3}). The 10d lift of
this is \bea
ds_{10}^2 &=& W^{1/2} H^{1/3}[ ds^2(AdS_3) + ds^2(M_7)],\\
ds^2(M_7) &=& \frac{dx^2}{4g^2 P(x)}+ \frac{4g^2P(x)}{C^2H(x)}
(dx^2)^2 \nonumber \\ &+& \frac{1}{g^2 W H^{2/3}}\sum_{I=1}^3
(x+3K_I) \left[ d\mu_I^2 + \mu_I^2\left( d\varphi_I +
\frac{g(x-\alpha_0)}{x+3K_I} dx^2 \right)^2 \right] \eea where the
$AdS_3$ satisfies $R_{\alpha \beta}= -(C^2/2)g_{\alpha\beta}$. This
solution was also encountered in~\cite{bubble} and studied in the
context of warped $AdS_3$ geometries. We should emphasise that any
warped $AdS_3$ geometry, where the internal space is compact, can be
thought of as a near-horizon geometry as can be seen by writing
$AdS_3$ as in (\ref{ads3}).  Therefore we will now perform a global
analysis of the above solution under the assumption that $M_7$ is
compact. The resulting horizon is this case will be topologically $S^1 \times M_7$. Much of the analysis parallels that in~\cite{CLPP}. The
metric on $M_7$ has a local $U(1)^4$ isometry. We will demand that
$M_7$ is a compact manifold, and thus since we require the metric on
$M_7$ to be globally regular we must have a global $U(1)^4$
symmetry. Examining scalar invariants constructed from the global
$U(1)^4$ symmetry generators implies $x$ is a globally defined
function on $M_7$. Therefore as in 5d we must take $x_1 \leq x \leq
x_2$ with $0<x_1<x_2<x_3$ roots of $P(x)$. This metric on constant
$x$ slices is positive definite everywhere except at the points
$x=x_1,x_2$ or $\mu_I=0$ where it degenerates. The latter points are
where $\partial / \partial \varphi_I$ vanish (as usual for an $S^5$)
and the corresponding conical singularities can be removed by
identifying $\varphi_I \sim \varphi_I +2 \pi$ as usual. Note that
$\mu_I$ are globally defined functions since they are the norms of
the globally defined $\partial / \partial \varphi_I$ Killing
vectors. The points $x=x_1,x_2$ were also degenerate points in 5d.
However the Killing vector which vanishes at these points in 10d is
modified: instead \be
V_i=\frac{k_i}{2g}\left(\frac{\partial}{\partial x^2} -
g(x_i-\alpha_0) \sum_{I=1}^3 \frac{1}{x_i+3K_I}  \frac{
\partial}{\partial \varphi_I} \right) \ee vanishes at $x=x_i$ where
$i=1,2$, where $k_i$ are constants. We see that these two Killing
vectors are distinct: this is qualitatively different to what
happens in 5d where it was the same Killing vector that vanished at
these two points ($ \partial /
\partial x^2$). In order to obtain a smooth metric these must
generate rotational symmetries and thus near $x=x_i$ one can
introduce coordinates $\chi_i \sim \chi_i +2\pi$ such that
$V_i=\partial / \partial \chi_i$. This is equivalent to removing the
conical singularities at $x=x_i$ and will fix the constants $k_i$.
For clarity we also introduce new coordinates $\Phi_I\sim
\Phi_I+2\pi$ such that $\partial/ \partial \Phi_I = \partial /
\partial \varphi_I$. The corresponding coordinate transformation
appropriate to the degeneration points $x=x_i$ are $(x^2, \varphi_I)
\to (\chi_i, \Phi_I)$ and defined by: \be \label{chiPhi} x^2 =
\frac{k_i}{2g} \chi_i, \qquad \varphi_I= \Phi_I -  C^I_i \chi_i \ee
where \be \label{C} C_i^I =\frac{(x_i-\alpha_0)k_i}{2(x_i+3K_I)}.
\ee  The absence of conical singularities at $x=x_i$ determines
$k_i^2$ and thus $k_i$ up to a sign. Without loss of generality we
choose signs such that \be \label{k} k_i= \frac{C^2
(x_i-\alpha_0)}{2g^2P'(x_i)}. \ee Note that $k_i$ and $C^I_i$ are
non vanishing since $H(x_i)>0$. One can verify that: \be
\label{sumC}\sum_{I=1}^3 C^I_i=1+k_i \ee

Since the {\it five} Killing vectors $V_i, \partial /
\partial \varphi_I$ span a {\it four} dimensional vector space with
basis $\partial /
\partial x^2, \partial / \partial \varphi_I$ they must be
linearly dependent. In order to avoid identifying arbitrarily close
points the dependency must take the form: \be \label{lindep}
\sum_{i=1}^2 n_i V_i + \sum_{I=1}^3 m_I \frac{\partial}{\partial
\varphi_I}=0\ee for integers $n_i,m_I$ which we may assume to be
coprime. It is easy to see that any subset of four of these five
integers must also be coprime. Therefore we must consider the
constraint on the parameters of the solution imposed by
(\ref{lindep}); this is: \be \label{DC} \sum_{i=1}^2 k_in_i =0,
\qquad m_I = \sum_{i=1}^2 C_i^In_i. \ee Note that (\ref{sumC}) then
implies \be \sum_{I=1}^3 m_I=n_1+n_2, \ee which allows us to write
$m_3$ in terms of the four other integers. Therefore in order to
obtain a smooth metric in 10d, the parameters of the solution
$(C^2,\alpha_0,K_I)$ need to be chosen such that (\ref{DC}) is
satisfied. In principle one can invert these relations to obtain a
metric parameterised by four integers $n_1,n_2,m_1,m_2$. In general,
however, it seems possible only to find expressions in terms of
these integers implicitly.

A special case in which it is straightforward to give explicit
formulas for these inversions, is $K_I=0$ for $I=1,2,3$ which
corresponds to the $AdS_3$ geometry of~\cite{ads3}. In this case we
have $3m=n_1+n_2$. For simplicity we use the scaling symmetry to set
$C=2g$ and thus $P(x)=x^3-(x-\alpha_0)^2$. In~\cite{KLR2} it was
shown that the existence of three positive distinct roots of $P$ is
equivalent to $0<\alpha_0<4/27$ and that we may introduce the
parameter $b$, which must satisfy $0<b<1$, such that \be \alpha_0=
\frac{4(1-b^2)^2}{(b^2+3)^3}, \qquad x_1= \frac{\alpha_0(b^2+3)}{4}
\qquad x_2= \frac{\alpha_0(b^2+3)}{(1+b)^2}, \qquad x_3=
\frac{\alpha_0(b^2+3)}{(1-b)^2}. \ee Therefore: \be k_1=
\frac{2(b^2+3)}{(b+3)(b-3)}, \qquad k_2= -\frac{b^2+3}{b(b+3)}. \ee
The condition $n_1k_1+n_2k_2=0$ can then be solved to give: \be
b=\frac{3n_2}{n_2-2n_1}. \ee This implies\begin{equation} \alpha_0 =
\frac{(2n_2-n_1)^2(n_2-2n_1)^2(n_1+n_2)^2}{27(n_1^2+n_2^2-n_2n_1)^3}
\end{equation}
The fact that $0<b<1$ is equivalent to either: (i) $n_1+n_2>0$ and
$n_1>0, n_2<0$, or (ii) $n_1+n_2<0$ and $n_1<0,n_2>0$. For case (i)
observe that defining $q=(n_1+n_2)/3=m$ and $p=-n_2$ gives the
positive integers $p,q$ of~\cite{ads3}. For case (ii)
$q=(n_1+n_2)/3$ and $p=-n_1$.

We now turn to the five-form and examine its regularity in the general case. We must prove this is regular
everywhere on $M_7$ and thus globally defined. Observe that $dx
\wedge dx^2$ is a globally defined non-vanishing 2-form, since near
the degeneration points $x=x_0,x_1$ this is proportional to the
volume form of the associated local $R^2$. Therefore $\epsilon_5=
H^{1/3}C^{-1} \textrm{vol}(AdS_3) \wedge dx \wedge dx^2$ is also
globally defined, which takes care of the first term in
(\ref{fiveform}). The second term in (\ref{fiveform}) contains
$\star_5 dX^I$ and thus we need to consider $\star_5dx$. One finds
that $\star_5 dx= 4g^2C^{-1}P(x) dx^2 \wedge \textrm{vol}(AdS_3)$ is
also globally defined, since $P(x)dx^2$ is regular at the
degeneration points $x=x_0,x_1$ (as can be seen in Cartesian
coordinates on the associated $R^2$). For the final term in
(\ref{fiveform}) we need to compute $\star_5F^I$; this is globally
defined since it looks like $f(x) \textrm{vol}(AdS_3)$ for some
regular function $f(x)$. Also note that as $x \to x_i$: \be
d\varphi_I+gA^I \sim d\Phi_I+\frac{k_i}{2(x_i+3K_I)} (x-x_i)d\chi_i
\ee and therefore, since $(x-x_i) d\chi_i$ is regular at $x=x_i$, we
see that $d\mu_I^2 \wedge (d\varphi_I+gA^I)$ must be globally
defined. This completes the proof that $F_5$ is regular everywhere
on $M_7$ and thus globally defined.

Note that in order to ensure this is a good solution of string
theory one must also show that the five form is appropriately
quantized. We will not consider this here, as we are only concerned
with proving existence of regular near-horizon geometries. As in the $K_I=0$ case~\cite{ads3}, however, we do not anticipate this will lead to further constraints on the integers $n_1,n_2,m_1,m_2$.


\begin{thebibliography}{99}

\bibitem{GR1}
  J.~B.~Gutowski and H.~S.~Reall,
  ``Supersymmetric AdS(5) black holes,''
  JHEP {\bf 0402}, 006 (2004)
  [arXiv:hep-th/0401042].

\bibitem{GR2}
  J.~B.~Gutowski and H.~S.~Reall,
  ``General supersymmetric AdS(5) black holes,''
  JHEP {\bf 0404}, 048 (2004)
  [arXiv:hep-th/0401129].

\bibitem{Chongetal}
  Z.~W.~Chong, M.~Cvetic, H.~Lu and C.~N.~Pope,
  ``General non-extremal rotating black holes in minimal five-dimensional
  gauged supergravity,'
  Phys.\ Rev.\ Lett.\  {\bf 95} (2005) 161301
  [arXiv:hep-th/0506029]

\bibitem{KLR}
  H.~K.~Kunduri, J.~Lucietti and H.~S.~Reall,
  `Supersymmetric multi-charge AdS(5) black holes,'
  JHEP {\bf 0604} (2006) 036
  [arXiv:hep-th/0601156].

\bibitem{adscft}
J.~M.~Maldacena, ``The large N limit of superconformal field
theories and supergravity,'' Adv.\ Theor.\ Math.\ Phys.\  {\bf 2},
231 (1998) [Int.\ J.\ Theor.\ Phys.\  {\bf 38}, 1113 (1999)]
[arXiv:hep-th/9711200],
S.~S.~Gubser, I.~R.~Klebanov and A.~M.~Polyakov, ``Gauge theory
correlators from non-critical string theory,'' Phys.\ Lett.\ B
{\bf 428}, 105 (1998) [arXiv:hep-th/9802109],
E.~Witten, ``Anti-de Sitter space and holography,'' Adv.\ Theor.\
Math.\ Phys.\  {\bf 2}, 253 (1998) [arXiv:hep-th/9802150].

\bibitem{KMMR}
  J.~Kinney, J.~M.~Maldacena, S.~Minwalla and S.~Raju,
  ``An index for 4 dimensional super conformal theories,''
  [arXiv:hep-th/0510251].

\bibitem{berkooz}
M.~Berkooz, D.~Reichmann and J.~Simon,
  ``A Fermi surface model for large supersymmetric AdS(5) black holes,''
  [arXiv:hep-th/0604023].

\bibitem{vacring}
  R.~Emparan and H.~S.~Reall,
  ``A rotating black ring in five dimensions,''
  Phys.\ Rev.\ Lett.\  {\bf 88}, 101101 (2002)
  [arXiv:hep-th/0110260].

\bibitem{GG}
  J.~P.~Gauntlett and J.~B.~Gutowski,
  ``All supersymmetric solutions of minimal gauged supergravity in five
  dimensions,''
  Phys.\ Rev.\  D {\bf 68} (2003) 105009
  [Erratum-ibid.\  D {\bf 70} (2004) 089901]
  [arXiv:hep-th/0304064].

\bibitem{KLR2}
H.~K.~Kunduri, J.~Lucietti and H.~S.~Reall,
  ``Do supersymmetric anti-de Sitter black rings exist?,''
  JHEP {\bf 0702} (2007) 026
  [arXiv:hep-th/0611351].


\bibitem{LLM}
  H.~Lin, O.~Lunin and J.~M.~Maldacena,
  ``Bubbling AdS space and 1/2 BPS geometries,''
  JHEP {\bf 0410} (2004) 025
  [arXiv:hep-th/0409174].

\bibitem{Chen}
  B.~Chen {\it et al.},
  ``Bubbling AdS and droplet descriptions of BPS geometries in IIB
  supergravity,''
  arXiv:0704.2233 [hep-th].

\bibitem{reduction}
  M.~Cvetic, H.~Lu, C.~N.~Pope, A.~Sadrzadeh and T.~A.~Tran,
  ``Consistent SO(6) reduction of type IIB supergravity on S(5),''
  Nucl.\ Phys.\  B {\bf 586} (2000) 275
  [arXiv:hep-th/0003103].

\bibitem{HDBH}
  H.~S.~Reall,
 ``Higher dimensional black holes and supersymmetry,''
  Phys.\ Rev.\ D {\bf 68} (2003) 024024
  [Erratum-ibid.\ D {\bf 70} (2004) 089902]
  [arXiv:hep-th/0211290].


\bibitem{KLR3}
  H.~K.~Kunduri, J.~Lucietti and H.~S.~Reall,
  ``Near-horizon symmetries of extremal black holes,''
  Class.\ Quant.\ Grav.\  {\bf 24} (2007) 4169
  [arXiv:0705.4214 [hep-th]].

\bibitem{SK}
  D.~Klemm and W.~A.~Sabra,
  ``General (anti-)de Sitter black holes in five dimensions,''
  JHEP {\bf 0102} (2001) 031
  [arXiv:hep-th/0011016].


\bibitem{GS}
  J.~B.~Gutowski and W.~Sabra,
  ``General supersymmetric solutions of five-dimensional supergravity,''
  JHEP {\bf 0510}, 039 (2005)
  [arXiv:hep-th/0505185].


\bibitem{Gut}
  J.~B.~Gutowski,
  ``Uniqueness of five-dimensional supersymmetric black holes,''
  JHEP {\bf 0408}, 049 (2004)
  [arXiv:hep-th/0404079].

\bibitem{Klemm}
  D.~Klemm and W.~A.~Sabra,
  ``Supersymmetry of black strings in D = 5 gauged supergravities,''
  Phys.\ Rev.\  D {\bf 62}, 024003 (2000)
  [arXiv:hep-th/0001131].

\bibitem{Cacciatori}
  S.~L.~Cacciatori, D.~Klemm and W.~A.~Sabra,
  ``Supersymmetric domain walls and strings in D = 5 gauged supergravity
  coupled to vector multiplets,''
  JHEP {\bf 0303} (2003) 023
  [arXiv:hep-th/0302218].

\bibitem{multibpsring}
  H.~Elvang, R.~Emparan, D.~Mateos and H.~S.~Reall,
  ``Supersymmetric black rings and three-charge supertubes,''
  Phys.\ Rev.\  D {\bf 71}, 024033 (2005)
  [arXiv:hep-th/0408120].

\bibitem{davis}
  P.~Davis, H.~K.~Kunduri and J.~Lucietti,
  ``Special symmetries of the charged Kerr-AdS black hole of D = 5 minimal
  gauged supergravity,''
  Phys.\ Lett.\  B {\bf 628} (2005) 275
  [arXiv:hep-th/0508169].

\bibitem{sinha1}
  A.~Sinha, J.~Sonner and N.~V.~Suryanarayana,
  ``At the horizon of a supersymmetric AdS(5) black hole: Isometries and
  half-BPS giants,''
  JHEP {\bf 0701} (2007) 087
  [arXiv:hep-th/0610002].

\bibitem{sinha2}
  A.~Sinha and J.~Sonner,
  ``Black Hole Giants,''
  arXiv:0705.0373 [hep-th].

\bibitem{gowdy}
  R.~H.~Gowdy,
  ``Vacuum space-times with two parameter spacelike isometry groups and compact
  invariant hypersurfaces: Topologies and boundary conditions,''
  Annals Phys.\  {\bf 83} (1974) 203.

\bibitem{Cvetic}
  M.~Cvetic {\it et al.},
  ``Embedding AdS black holes in ten and eleven dimensions,''
  Nucl.\ Phys.\  B {\bf 558} (1999) 96
  [arXiv:hep-th/9903214].

\bibitem{bubble}
  J.~P.~Gauntlett, N.~Kim and D.~Waldram,
  ``Supersymmetric AdS(3), AdS(2) and bubble solutions,''
  JHEP {\bf 0704} (2007) 005
  [arXiv:hep-th/0612253].


\bibitem{CLPP}
  M.~Cvetic, H.~Lu, D.~N.~Page and C.~N.~Pope,
  ``New Einstein-Sasaki spaces in five and higher dimensions,''
  Phys.\ Rev.\ Lett.\  {\bf 95} (2005) 071101
  [arXiv:hep-th/0504225].

\bibitem{ads3}
  J.~P.~Gauntlett, O.~A.~P.~Mac Conamhna, T.~Mateos and D.~Waldram,
  ``Supersymmetric AdS(3) solutions of type IIB supergravity,''
  Phys.\ Rev.\ Lett.\  {\bf 97} (2006) 171601
  [arXiv:hep-th/0606221].

\bibitem{rigidity}
  S.~Hollands, A.~Ishibashi and R.~M.~Wald,
  ``A higher dimensional stationary rotating black hole must be
  axisymmetric,''
  Commun.\ Math.\ Phys.\  {\bf 271} (2007) 699
  [arXiv:gr-qc/0605106].

\bibitem{half}
  J.~B.~Gutowski and W.~A.~Sabra,
  ``Half-Supersymmetric Solutions in Five-Dimensional Supergravity,''
  arXiv:0706.3147 [hep-th].


\bibitem{her}
 P.~Figueras, C.~A.~R.~Herdeiro and F.~P.~Correia,
 ``On a class of 4D Kahler bases and AdS(5) supersymmetric Black Holes''   [arXiv:hep-th/0608201].


\end{thebibliography}
\end{document}